\documentclass[a4paper,11pt]{article}
\pdfoutput=1
\usepackage{jcappub} 
\usepackage{natbib}
\usepackage{graphicx,color}
\usepackage[normalem]{ulem}

\usepackage[T1]{fontenc} 
\usepackage{lineno, booktabs}

\newcommand{\lsim}{\mathrel{\mathop{\kern 0pt \rlap
  {\raise.2ex\hbox{$<$}}}
  \lower.9ex\hbox{\kern-.190em $\sim$}}}
\newcommand{\gsim}{\mathrel{\mathop{\kern 0pt \rlap
  {\raise.2ex\hbox{$>$}}}
  \lower.9ex\hbox{\kern-.190em $\sim$}}}

\newcommand{\reply}[1]{\textcolor{black}{#1}} 
\newcommand{\replyy}[1]{\textcolor{black}{#1}}

\title{Multi-messenger constraints to the local emission of cosmic-ray electrons}

\author[a,b]{S. Manconi}
\author[c,d]{M. Di Mauro,}
\author[a,b]{F. Donato}

\affiliation[a]{Dipartimento di Fisica, Universit\`a di Torino, via P. Giuria 1, 10125 Torino, Italy}
\affiliation[b]{Istituto Nazionale di Fisica Nucleare, Sezione di Torino, Via P. Giuria 1, 10125 Torino, Italy}
\affiliation[c]{NASA Goddard Space Flight Center, Greenbelt, MD 20771, USA}
\affiliation[d]{Catholic University of America, Department of Physics, Washington DC 20064, USA}

\emailAdd{manconi@to.infn.it}
\emailAdd{mdimauro@slac.stanford.edu}
\emailAdd{donato@to.infn.it}

  \abstract{
The data on the inclusive flux of cosmic positrons and electrons ($e^++e^{-}$) have been recently collected from GeV to tens of TeV energies by several experiments with unprecedented precision. In addition, the {\it Fermi}-LAT Collaboration has provided a new energy spectrum for the upper bounds on the $e^++e^{-}$ dipole anisotropy. 
This observable can bring information on the emission from local Galactic sources, notably measured with high precision at radio frequencies.
 We develop a framework in which $e^-$  and $e^+$ measured at Earth from GeV up to tens of TeV energies have a composite origin.   
A dedicated analysis is deserved to Vela YZ and Cygnus Loop Supernova Remnants (SNRs), 
\reply{for which we consider two different models for the injection of $e^-$.}
We investigate the consistency of these models using the three physical observables: the {\it radio flux} from  Vela YZ and Cygnus Loop at all the available frequencies, 
 the {\it $e^++e^-$ flux} from five experiments from the GeV to tens of TeV energy, 
 the {\it  $e^++e^-$ dipole anisotropy} upper limits from 50 GeV to about 1 TeV.
 We find that the radio flux for these nearby SNRs strongly constraints the properties of the injection electron spectrum, 
 partially compatible with the looser constraints derived from the   $e^+ + e^-$ flux data. We also  perform a multi-wavelength multi-messenger analysis by fitting
 simultaneously the radio flux on Vela YZ and Cygnus Loop and the $e^+ + e^-$ flux, and checking the outputs against the  $e^+ + e^-$ dipole anisotropy data. 
Remarkably, we find a model which is compatible with all the $e^++e^-$ flux data, the radio data for Vela YZ and Cygnus Loop, and with the anisotropy upper bounds. 
We show the severe constraints imposed by the most recent data on the  $e^+ + e^-$ dipole anisotropy.   
}

\begin{document}
\maketitle
\flushbottom

\section{Introduction}
\label{sec:intro}
The {\it flux} of cosmic-ray (CR) electrons and positrons ($e^-$ and $e^+$) has been measured with unprecedented precision over more than four orders of magnitude of energy. 
One of the most accurate measurements on single CR $e^-$ and $e^+$ and inclusive ($e^+ + e^-$) fluxes is provided by 
AMS-02 on board the International Space Station (ISS), between 0.1 GeV to 1 TeV energy, and with errors reaching the few percent level 
\citep{2014PhRvL.113l1101A,2014PhRvL.113l1102A,Aguilar:2014fea}. 
The {\it Fermi} Large Area Telescope (LAT) has collected  almost seven years of $e^+ + e^-$ events in the 7~GeV-2~TeV energy range \citep{2017PhRvD..95h2007A}.
CALET on the ISS, and HESS on the ground, are providing $e^+ + e^-$ data up to  3 TeV and 30 TeV energy, respectively \citep{PhysRevLett.119.181101, HESSICRC17, PhysRevLett.120.261102}.
The DAMPE Collaboration has recently reported the direct detection of a break at around 1 TeV in the flux of the $e^+ + e^-$ measured between 25 GeV to 4.6 TeV \citep{Ambrosi:2017wek}.
Many theoretical interpretations have been proposed for the AMS-02 lepton data, invoking sources of  $e^{+}$ and/or $e^-$ in the Interstellar Medium (ISM), 
from Pulsar Wind Nebulae (PWNe) and  Supernova Remnants (SNRs)\cite{Kobayashi:2003kp,Pohl:2012xs, 2009APh....32..140G, Gaggero:2013nfa,Delahaye:2014osa,DiMauro:2014iia,Manconi:2016byt}, and also 
 in the context of annihilation and decay of dark matter particles \cite{Bergstrom:2013jra,DiMauro:2015jxa}. 
In addition to the flux, the LAT team has also published the spectrum of  upper limits on the  $e^+ + e^-$ {\it dipole anisotropy} \cite{Abdollahi:2017kyf}.
Since the typical propagation length of TeV $e^\pm$  is smaller than $\sim 0.3$~kpc, $e^+$ and $e^-$ detected at TeV energies are most probably emitted from local sources, leaving a possible signature in the dipole anisotropy \cite{Manconi:2016byt}.
At variance, nuclei suffer mainly from diffusion rather than energy losses, so the hadronic flux from local sources is typically spread, and sets below the cumulative contribution of all Galactic sources.

The  contribution from the local source candidates is usually associated with high uncertainties, primarily connected to the properties of the accelerated and emitted $e^-$ and $e^+$.
Moreover, the completeness of current catalogs, such as SNRs, is assessed by means of the observed surface brightness (see e.g. \cite{2015MNRAS.454.1517G}), thus leaving open the possibility that nearby and 
very old sources may contribute to the flux at Earth even if they are no longer visible at any energy of the electromagnetic band.
A strategy to constrain the source contributions of local known sources is to model their multi-wavelength emission and to connect it to the emitted CRs.
For example, the lepton emission from sources embedded in a magnetic field, such as $e^-$ from SNRs, can be connected with their synchrotron emission at {\it radio} frequencies 
(see \cite{2010A&A...524A..51D,DiMauro:2014iia,Manconi:2016byt} and references therein). 
In addition, the most recent experimental upper bounds on the dipole anisotropy could set further limits on the properties of local and dominant sources.

In the present paper we use this strategy to quantify the contribution of local known sources, in particular from two SNRs which are widely considered as the main candidates to contribute significantly to the high energy part of the $ e^-$ flux at Earth
(often measured cumulatively through $e^-+e^+$), namely Vela and Cygnus Loop, see e.g. \cite{Kobayashi:2003kp}. 
For the first time, we present a multi-component model that explains the $e^+$ and $e^-+e^+$ fluxes from five experiments and in a wide energy range, and that is simultaneously compatible with the upper bounds on the dipole anisotropy and the radio emission from the most intense and closest SNRs.
The paper is structured as follows.  
\reply{Our model for the cosmic-ray electrons from SNRs is outlined in Sec.~\ref{sec:model}.} In Sec.~\ref{sec:radio} the constrains imposed by radio data on our sample of local SNRs are presented. The constraints imposed from $e^-+e^+$ and dipole data are discussed respectively in Sec.~\ref{sec:flux} and Sec.~\ref{sec:dipole}. A model combining the multi-wavelength data for local SNRs that explains the most recent flux and dipole data is presented in Sec.~\ref{sec:multiw}, before concluding in Sec.~\ref{sec:conc}.

\section{\label{sec:model}Cosmic-ray electrons from SNRs}
\reply{Cosmic-ray $e^\pm$} can be injected \reply{in the interstellar medium (ISM)} by shocked stellar 
environments - SNRs as well as PWNe -  according to the first order Fermi acceleration mechanism
(for a comprehensive review on the SNR paradigm for Galactic CRs see \cite{Blasi:2013rva} and references therein).
\reply{We  focus here on SNRs.}
For a detailed treatment of the injection of $e^-$ by SNRs and their propagation in the Galaxy we refer to \cite{2010A&A...524A..51D, Manconi:2016byt}.
\reply{We here remind the basics of our model, along with an additional new treatment for the injection of $e^-$ by SNRs and for the synchrotron radio emission from known SNRs. }

\subsection{Injection of cosmic-ray electrons from SNRs into the ISM}
The details of the release mechanism of $e^-$ from SNRs are poorly known and still under debate \cite{2010APh....33..160C, 2012MNRAS.427...91O, Blasi:2013rva, 2009MNRAS.396.1629G}, and could affect the properties of the escaping $e^-$, above all the energy spectrum. 
\reply{We implement here two different models, the burst-like injection and the evolutionary model. }

\reply{The injection of $e^-$ accelerated by SNRs is commonly described  through a \textit{burst-like} approximation \cite{2010A&A...524A..51D}, 
in which all the $e^-$ are released in the ISM at a time equal to the age of the source. Under this hypothesis,} the energy spectrum $Q(E)$ of accelerated $e^-$ can be described by the function
 \begin{equation}
 Q(E)= Q_{0} \left( \frac{E}{E_0}\right)^{- \gamma} \exp \left(-\frac{E}{E_c} \right), 
 \label{eq:Q_E}
\end{equation}
 where $Q_{0}$ is in units of GeV$^{-1}$, $E_c$ is a cutoff energy and $E_0= 1$~GeV.
Given the injection spectrum in Eq.~\ref{eq:Q_E}, the total energy emitted in $e^-$ from SNR (or $e^\pm$ for PWN) in units of GeV (or erg) can be obtained as (see \cite{2010A&A...524A..51D})
\begin{equation}
 E_{\rm tot} = \int _{E_1} ^\infty dE \, E \,Q(E) \,,
 \label{eq:Etot}
\end{equation}
where we fix $E_1 = 0.1$ GeV. The normalization of the spectrum in Eq.~\ref{eq:Q_E} can be constrained 
from available catalog quantities for single sources \reply{(see Sec.~\ref{sec:synchro})}, or by using average population properties for the smooth galactic component   \cite{2010A&A...524A..51D,DiMauro:2014iia,Manconi:2016byt}. 
\reply{The burst-like assumption is considered appropriate at high energy, since $e^-$ of energy $E\gsim100$ GeV are believed to be released  within a few kyr from the initial burst, and this timescale is much smaller than the age of the sources typically considered  to explain the CR $e^-$ data at Earth \cite{2009MNRAS.396.1629G}.}

\reply{In addition to the burst-like approximation, we also implement the {\em evolutionary model}  for the escape  of $e^-$ from SNRs as derived in Ref.~\cite{2012MNRAS.427...91O}.
The authors of Ref.~\cite{2012MNRAS.427...91O} assume analytical models for the temporal evolution of the shock radius and its velocity. 
They also derive the timescales and the space-energy distributions for trapped and runaway $e^-$. 
During the Sedov phase, the escape-limited maximum energy $E_{\rm m, esc}(T)$ below which CRs  are still trapped in the SNR is defined as \cite{2012MNRAS.427...91O}:
\begin{equation}\label{eq:Emesc}
 E_{\rm m, esc}(t)= E_{\rm knee} \left(\frac{T}{t_{\rm Sedov}}\right)^{-\alpha}
\end{equation}
where $E_{\rm knee}=10^{15.5}$~eV is the energy at the knee of the CR all-particle spectrum, $t_{\rm Sedov}$  is the start time of the Sedov phase, $\alpha$ describes the evolution of the maximum energy during the Sedov phase, and $T$ is the SNR age. 
For energies smaller than $E_{\rm m, esc}(T)$, the $e^-$ are still trapped in the SNR, and their energy spectrum is described as:
\begin{align}\label{eq:QTrap}
  Q_{\rm trap}(E, T)&= A \left( \frac{E_{\rm m, esc}(T)}{E_0}\right)^{- (\gamma + \beta/\alpha)} \left(\frac{E}{E_{\rm m, esc}(T)}\right)^{-\gamma} \exp \left(-\frac{E}{E_c} \right) = \\
  &=  Q_{\rm 0, trap}(T) \left(\frac{E}{E_{0}}\right)^{-\gamma} \exp \left(-\frac{E}{E_c} \right) \label{eq:QTrap2}
\end{align}
where $A$ is a normalization factor, and $\beta$ describes the evolution of the  electron number inside the SNR. 
The $Q_{\rm 0, trap}(T)$ is obtained by recasting Eq.~\ref{eq:QTrap} using the explicit form of $E_{\rm m, esc}(T)$ in Eq.~\ref{eq:Emesc}.  
The energy spectrum $Q_{\rm esc}(E)$ of runaway $e^-$ is instead described by:
\begin{equation}\label{eq:Qesc}
  Q_{\rm esc}(E)= A \left( \frac{E}{E_0}\right)^{- (\gamma + \beta/\alpha)} \exp \left(-\frac{E}{E_c} \right)\quad. 
\end{equation}
The  $e^-$ at a given  energy $E$ escape from the SNR at a time:
\begin{equation}\label{eq:Tesc}
 T_{\rm esc}=t_{\rm Sedov} \left( \frac{E}{E_{\rm knee}}\right)^{-\frac{1}{\alpha}}\quad. 
\end{equation}}
\reply{The idea of an escape-limited maximum energy $E_{\rm m, esc}(T)$, or equivalently of $T_{\rm esc}$ after which $e^-$ of a given energy can run away from the SNR, 
 determines the difference between the simpler burst-like approximation and this evolutionary escape model. 
Focusing on specific sources, as for example the Vela YZ and Cygnus Loop SNRs ($T=$11.3~kyr and 20~kyr respectively, see below), 
this escape model states that CRs $e^-$ with energies  $E<E_{\rm m, esc}=88$~GeV  and  $E<E_{\rm m, esc}=17$~GeV  are still trapped in Vela YZ and Cygnus Loop, respectively. 
On the opposite,  $e^-$ with energies $E>E_{\rm m, esc}$ have been released in the ISM. 
Also, we note that the energy spectral index of runaway $e^-$ is modified with respect to the one of trapped ones, as stated by Eqs. \ref{eq:QTrap} and \ref{eq:Qesc}.  
 We will study the consequences of these spectral modifications in the following Sections. 
The specific values of $E_{\rm m, esc}(T)$, or equivalently the ages of the two sources, make the burst like approximations a more suitable description for the older Cygnus Loop than for Vela YZ. 
 We fix $t_{\rm Sedov}=200$~yr, $\alpha=2.6$ and $\beta=0.6$, as in Ref.~\cite{2012MNRAS.427...91O}. 
When not differently stated, our results are shown for the burst-like model. 
}

\subsection{Propagation in the Galaxy}
The propagation of  $e^{-}$ and $e^{+}$ from their sources to the Earth has been treated as in \cite{2010A&A...524A..51D,DiMauro:2014iia,Manconi:2016byt}, 
\reply{from which we remind here some basic ingredients. We refer the reader to Ref.~\cite{Manconi:2016byt} for futher details.
The cosmic-ray $e^{-}$ and $e^{+}$ number density $\psi = \psi(E, \mathbf{x}, t)\equiv dn/dE$ per unit volume and energy 
obeys the  transport equation:
\begin{equation}
 \frac{\partial \psi}{\partial t}  - \mathbf{\nabla} \cdot \left\lbrace K(E)  \mathbf{\nabla} \psi \right\rbrace + 
 \frac{\partial }{\partial E} \left\lbrace \frac{dE}{dt} \psi \right\rbrace = q(E, \mathbf{x}, t)
 \label{eq:diff}
\end{equation}
where $K(E)$ is the energy dependent diffusion coefficient, $dE/dt\equiv b(E)$ accounts for the energy losses and $q(E, \mathbf{x}, t)$ is the  $e^-$ and $e^+$ source term. The flux of electron $\Phi$ at the Earth is connected to the number density through $\Phi=v /4\pi \;\psi$. 
}
We solve the transport equation in Eq.~\ref{eq:diff} in a semi-analytic model,  
assuming a spatially uniform diffusion coefficient: 
\begin{equation}
\label{eq:diff_coeff}
 K(E)= \beta K_0 (\mathcal{R}/1 \text{GV})^\delta \simeq K_0 (E/1 \text{GeV})^{\delta}
\end{equation}
where $\beta=v/c$ (for relativistic $e^\pm$, as in this analysis, $\beta=1$)
and $\mathcal{R}$ is the particle rigidity. 
We include $e^\pm$ energy losses by Inverse Compton  scattering off the interstellar radiation field, 
and synchrotron losses on the Galactic magnetic field.
A full-relativistic treatment of Inverse Compton losses has been implemented in the Klein-Nishina regime, according to Ref.~\cite{2010A&A...524A..51D}. 
\reply{The black body approximation for the interstellar photon populations
at different wavelengths has been taken from \cite{2010A&A...524A..51D} (model M2 in their Table 2).
The Galactic magnetic field intensity 
has been assumed $B=3.6\; \mu$G, as resulting from the sum (in quadrature) of the regular and turbulent components \citep{2007A&A...463..993S}.} 
At the energies considered here,  the energy losses dominate over  diffusion effects.
Therefore, modifications of the diffusion coefficient are not expected to modify significantly our conclusions.
The propagation parameters are fixed according to the fits to CR data performed within a semi-analytical diffusion model in \cite{Kappl:2015bqa} (K15) and \cite{Genolini:2015cta} (G15) 
(see also \cite{Manconi:2016byt}). 
As for the K15 model, it is found to be $K_0=0.0967$ kpc$^2$/Myr and \reply{$\delta=0.408$, while for the G15 model $K_0=0.05$ kpc$^2$/Myr and $\delta=0.445$}.
The values found in these two papers are also compatible with the ones derived in \cite{Johannesson:2016rlh,Korsmeier:2016kha}.

\reply{Given our focus on single sources, we report here the explicit solutions of the time-dependent transport equation for the CR flux from a single source of $e^-$, which  can be  also found in several literature works (see  e.g. \cite{PhysRevD.52.3265,Aharonian:1995zz,Mlyshev:2009twa, 2010A&A...524A..51D}).  
In the burst-like approximation, the CR $e^-$ and $e^+$ density $\mathcal{\psi}(E,\mathbf{x})$ at a position $\mathbf{x}$ (in Galactic coordinates) and energy $E$, and   considering an infinite diffusion halo, reads:
 \begin{equation}\label{eq:singlesourcesolution}
  \mathcal{\psi}(E,\mathbf{x}) = \frac{b(E_s)}{b(E)} \frac{1}{(\pi \lambda^2)^{\frac{3}{2}}} \exp\left({-\frac{|\mathbf{x} -\mathbf{x_{s}} |^2}{ \lambda^2}}\right)Q(E_s)
\end{equation}
where $b(E)$ is the energy loss function, $\mathbf{x_{s}}$ indicates the source position. $\lambda$ is the typical propagation scale length:
\begin{equation}
\label{eq:lambda}
 \lambda^2= \lambda^2 (E, E_s) \equiv 4\int _{E} ^{E_s} dE' \frac{K(E')}{b(E')},
\end{equation} 
where \replyy{$E_s(E)\equiv E_s(E;t,t_s)$}  is the initial energy of $e^\pm$ that cool down to $E$ in a {\rm loss time} $\Delta \tau$:
\begin{equation}
 \Delta \tau (E, E_s) \equiv \int_{E} ^{E_s} \frac{dE'}{b(E')} = t-t_{{\rm s}} \quad , 
\end{equation} 
and $t_{{\rm s}}$ is the source age.
}
\reply{As for the evolutionary escape model in Ref.~\cite{2012MNRAS.427...91O} the $\mathcal{\psi_{\rm esc}}(E,\mathbf{x})$ of runaway CR $e^\pm$ at a position $\mathbf{x}$  and energy $E$, and   considering an infinite diffusion halo is:
 \begin{equation}\label{eq:sol_esc}
  \mathcal{\psi}_{\rm esc}(E,\mathbf{x}) = \frac{b(E_s(E))}{b(E)} \frac{1}{(\pi \lambda^2)^{\frac{3}{2}}} \exp\left({-\frac{|\mathbf{x} -\mathbf{x_{s}} |^2}{ \lambda^2}}\right)Q_{\rm esc}(E_s(E))
\end{equation}
where $E_s$ is again the initial energy of $e^\pm$ that cool down to $E$, defined now as:
\begin{equation}
 \int_{E} ^{E_s(E)} \frac{dE'}{b(E')} = t-T_{\rm esc}(E_s)\,,
\end{equation}
\replyy{and $T_{\rm esc}$ is given by Eq.~\ref{eq:Tesc}.}
\replyy{The energy spectrum $Q_{\rm esc}(E_s(E))$ is defined in Eq.~\ref{eq:Qesc}, and is obtained as \cite{2012MNRAS.427...91O}:
\begin{equation}
Q_{\rm esc}(E) = \int dt\int d \mathbf{x} \, q(E, \mathbf{x}, t)\,,
\end{equation}
and the solution reported in Eq.~\ref{eq:sol_esc} is given for a source term in the form $q(E, \mathbf{x}, t)= \delta(\mathbf{x}) \delta (t-T_{\rm esc}(E))Q_{\rm esc}(E) )$.}
}

\subsection{The radio synchrotron emission from nearby SNRs}
\label{sec:synchro}
One of the key points of this paper is the inspection of selected near SNRs in terms of the available flux radio data, in order to obtain a better understanding of  $e^-$ and $e^+$ flux data. 
The very-high-energy $e^-$ and $e^+$ flux and the radio data for nearby sources are connected under the hypothesis that the radio emission from the source is due to synchrotron radiation from $e^-$ accelerated and interacting with the SNR magnetic field.

Under the hypothesis that the radio emission from the source is due to synchrotron radiation from $e^-$ accelerated and interacting with the SNR magnetic field $B$, the normalization of the injection spectrum $Q_{0,\rm{SNR}}$ can be connected to the radio flux density $B^{\nu}_r(\nu)$:
\begin{equation}
 \label{eq:Br}
Q_{0,\rm{SNR}} = 1.2 \cdot 10^{47}  \text{GeV}^{-1} (0.79)^{\gamma}  \frac{B^{\nu}_r(\nu)}{\text{Jy}}
 \left[ \frac{d}{\text{kpc}}\right]^{2} 
\left[ \frac{\nu}{\text{GHz}}\right]^{\frac{\gamma -1}{2}}
\left[ \frac{B}{100 \mu\text{G}}\right]^{-\frac{\gamma+1}{2}}.
\end{equation}
The derivation of this expression is extensively provided in \cite{2010A&A...524A..51D}, and it was successively used also in \cite{DiMauro:2014iia,Manconi:2016byt}.
The energy emitted at a given frequency is radiated at the energy loss rate \replyy{$b(E(\nu))$}
(see Eqs.~48-50 in \cite{2010A&A...524A..51D}), and one implicitly assumes that observation takes place within the
time-interval $E(\nu)/b(E(\nu))$ \replyy{($E(\nu)=h \nu$)} during which the $e^-$ radiate
after the burst, and that the flux is quasi constant within this time interval.
We note in Eq.\eqref{eq:Br} the well-known relation between the index of the $e^-$ distribution $\gamma$ and the radio index $\alpha_r= (\gamma -1)/2$.
\reply{The $Q_{0,\rm{SNR}}$ term in Eq. \ref{eq:Br} is implemented by Eq. \ref{eq:Q_E} in case of the burst-like approximation. 
In case of the evolutionary escape model, we instead implement in Eq.~\ref{eq:Br} the $Q_{0,\rm{trap}}$ of Eq.~\ref{eq:QTrap2}. }
 \begin{figure}[t]
  \includegraphics[width=0.5\textwidth]{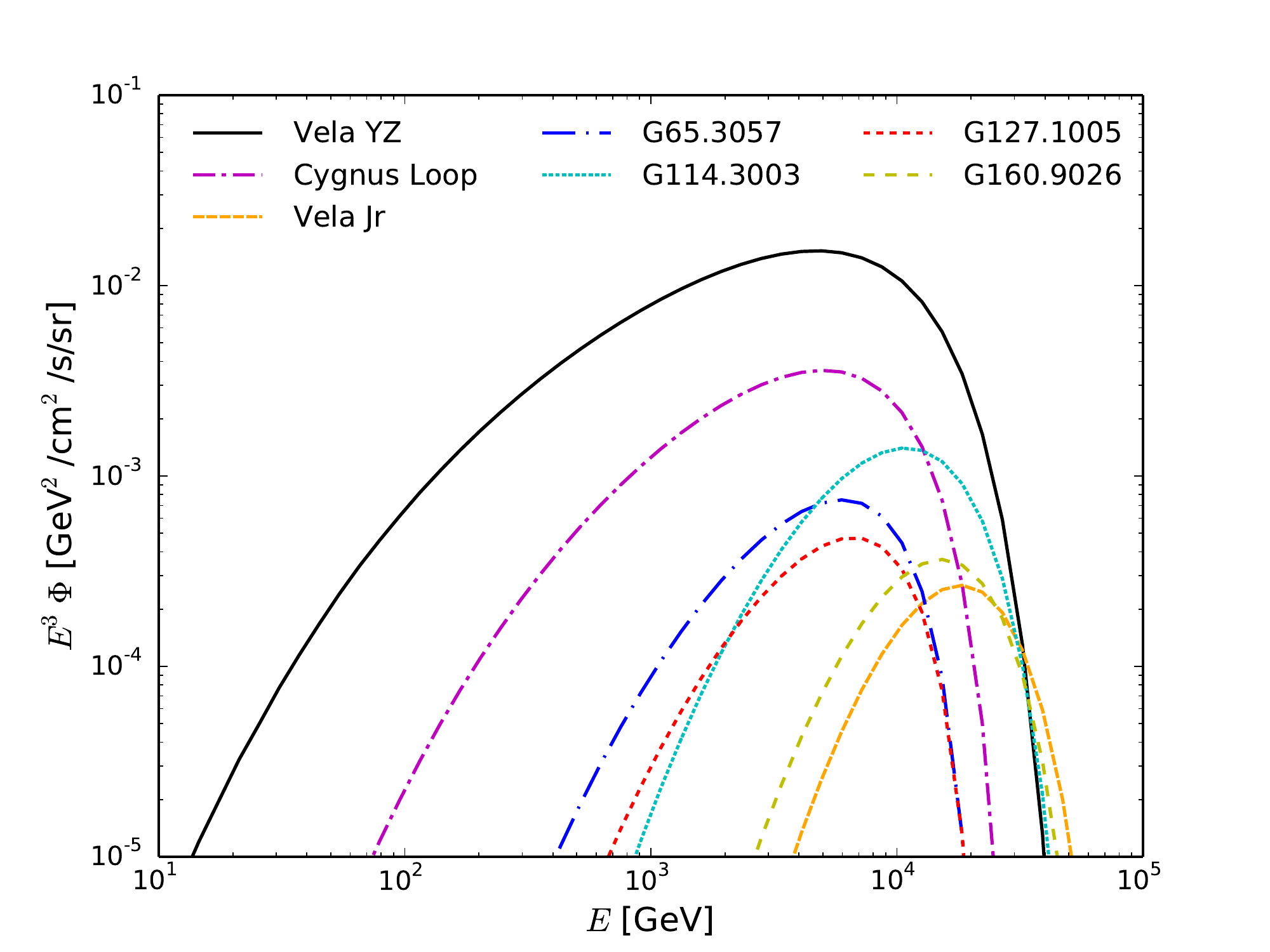}
  \includegraphics[width=0.5\textwidth]{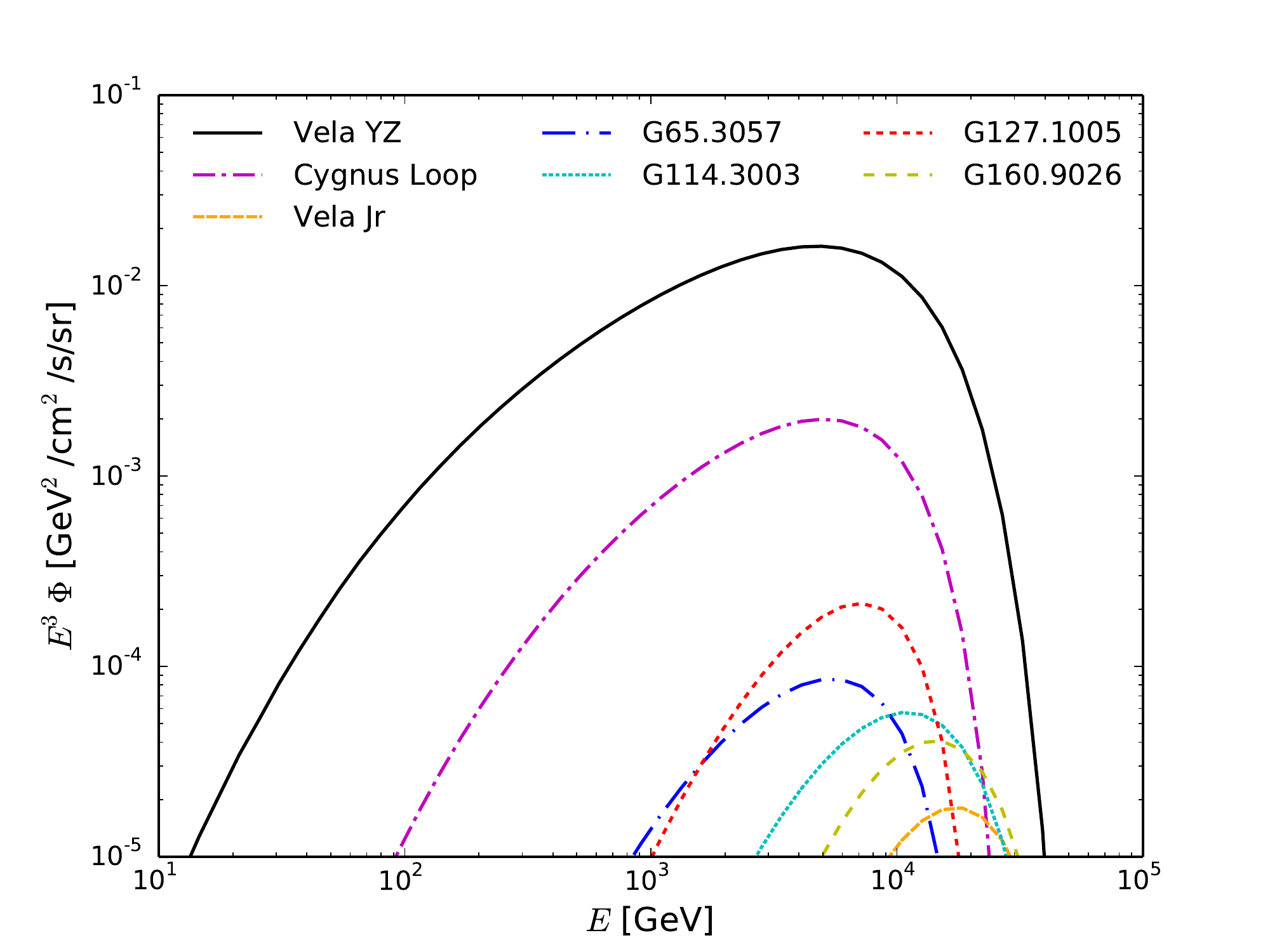} 
  \caption{Electron flux at Earth from near SNRs in the Green catalog at $d<1$~kpc from the Earth.
  Left: A common spectral index of $\gamma=2.0$ and a total energy released in $e^-$ of $E_{\rm tot}=7 \cdot 10^{47}$~erg has been assumed for each source. 
  Right: The spectral index and the $Q_0$ for each source are fixed  according to the catalog data and Eq.\ref{eq:Br} for a single frequency. 
  All the curves are computed for $E_c=10$~TeV and K15 propagation model. 
  }\label{fig:nearsnr}
 \end{figure} 

Our search for the sources that can contribute most to the  $e^-$ flux rests on the computation of the  $e^-$ from 
catalogued sources. We consider the sources in the Green SNR catalog \cite{Green:2014cea}, and  find seven SNRs which are located at $d<1$ kpc from the Earth. 
In order to illustrate the role of these SNRs,  we compute their flux of $e^-$ at Earth. We first assume that they all inject  $e^-$ in the ISM with the 
 common spectral index of $\gamma=2.0$ and with a total energy released in $e^-$ of $E_{\rm tot}=7 \cdot 10^{47}$~erg, as very often assumed in the literature (see e.g. \cite{Kobayashi:2003kp}). 
The only catalogued parameters here are the distance and the age of the source. 
The results are shown in the left panel of Fig.~\ref{fig:nearsnr}. Vela YZ  turns out to be the most powerful source, followed by Cygnus Loop. Electrons from the other sources have fluxes smaller than up one order of magnitude. 
Indeed, the Green  catalog \cite{Green:2014cea} also provides the spectral index and the radio properties for each source that, 
when implemented in Eq.~\ref{eq:Q_E}, lead to the fluxes in Fig.~\ref{fig:nearsnr}, right panel. 
This more realistic approach demonstrates that the only two powerful sources are indeed Vela YZ and Cygnus Loop, 
while the other SNRs contribute with an $e^-$ flux at Earth which is at the percent level of the Vela YZ one. 
We identify Vela YZ and Cygnus Loop  as the candidates 
expected to contribute most significantly to the high-energy tail of $e^{+}+e^{-}$ flux, given their distance, age and radio flux \cite{Kobayashi:2003kp,DiMauro:2014iia,Manconi:2016byt}. 
As shown in the following, Vela Jr can emerge as a significant contributor to the $e^{+}+e^{-}$ flux in the TeV range when the leptonic model inferred in \cite{2008ApJ...678L..35K} is considered, given the high value for the cutoff of $E_c=25$~TeV and the low magnetic field ($12\mu$G).


\section{\label{sec:radio}Results on the SNR  properties from radio data}
 \begin{figure}[t]
  \includegraphics[width=0.5\textwidth]{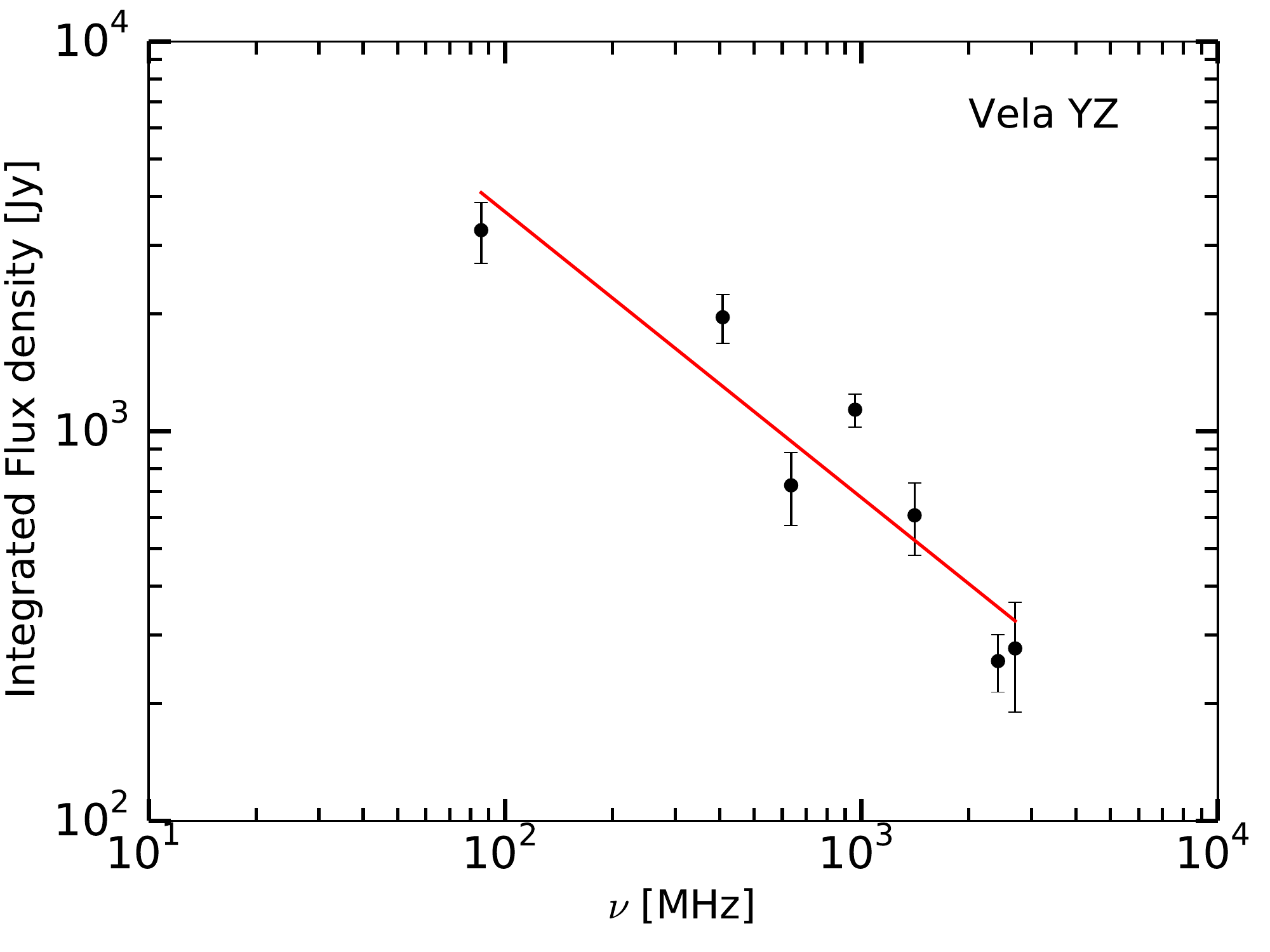}
  \includegraphics[width=0.5\textwidth]{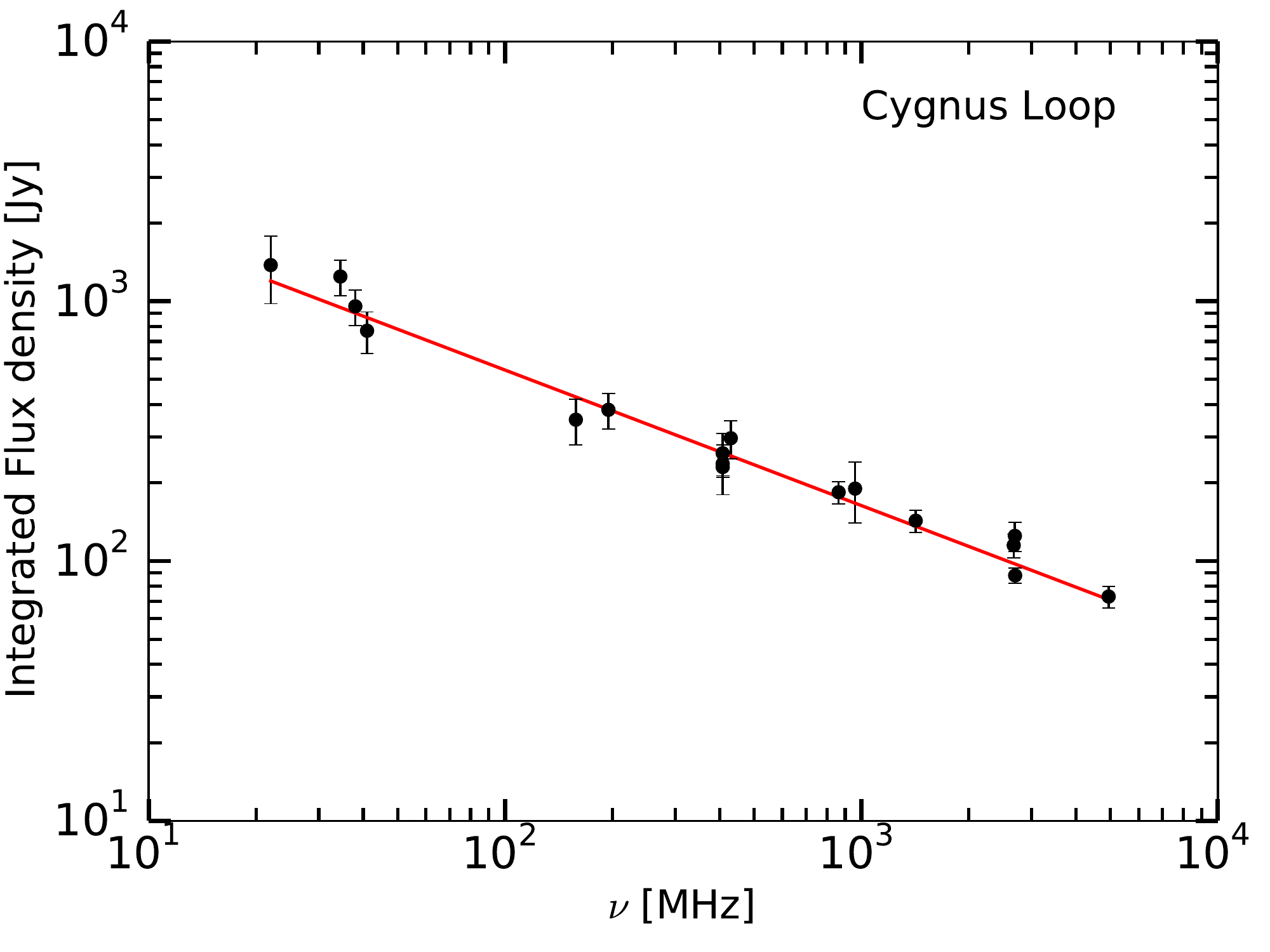} 
  \caption{A fit to the radio spectrum of Vela SNR (left panel) and Cygnus Loop (right panel) using Eq.~\eqref{eq:Br}. 
   The red line represents the best fit model to the data. 
   The integrated flux densities $B_r$ are taken from \cite{2001A&A...372..636A, 2004A&A...426..909U}.
  }\label{fig:radio}
 \end{figure} 
With respect to previous analysis where usually a single frequency was considered (see, e.g., \cite{DiMauro:2014iia,DiMauro:2017jpu}), 
we use here the radio spectrum in the widest available range of frequencies:
from 85.7~MHz to 2700~MHz for Vela YZ \cite{2001A&A...372..636A} and from $22$~MHz to $4940$~MHz for Cygnus Loop \cite{2004A&A...426..909U}.
We fix the Vela YZ (Cygnus Loop) distance and age to be: $d=$ 0.293~kpc (0.54~kpc) and $T=$11.3~kyr (20~kyr) \cite{2003ApJ...596.1137D,2005AJ....129.2268B,1994PASJ...46L.101M,2004A&A...426..909U}, respectively. 
The magnetic field of galactic SNRs is often inferred from multi-wavelength analysis, and the values typically range between few $\mu$G to even $10^3\mu$G \cite{2012SSRv..166..231R}. 
The magnetic field of Vela YZ is here fixed to $B=$ 36 $\mu G$, corresponding to a mean of the values inferred from X-ray data for the Y and Z regions \cite{Sushch:2013tna}, while for Cygnus Loop  we consider the best fit value of $B=60$~$\mu G$ of the hadronic model for the gamma-ray analysis in \cite{2011ApJ...741...44K}.
In Fig.~\ref{fig:radio}  we display the results for the fit to the available radio data of both Vela YZ and Cygnus Loop. 

We then invert Eq.~\ref{eq:Br} to fit $B^{\nu}_r(\nu)$ as a function of $\gamma$ and $Q_{0,\rm{SNR}}$ for all the 
available frequencies $\nu$. 
We tune the injection spectrum of local SNRs in order to reproduce the radio data, since at this wavelength the $e^-$ are the main emitters.
It is worth noting that \reply{in the case of burst-like approximation} we work under the assumption that the electromagnetic emission we observe today from those SNRs reflects the properties of the  $e^-$ population that has been released and injected in the ISM. 
The best fit parameters are: 
$\gamma_{\rm{Vela}}= 2.47\pm0.10$, $E_{\rm{tot}, \rm Vela}= (2.28\pm0.06)\cdot 10^{47}$ erg,  $\gamma_{\rm{Cygnus}}= 2.04\pm0.04 $ and $E_{\rm{tot}, \rm Cygnus}= (1.18\pm 0.16)\cdot 10^{47}$ erg. The numbers for the Vela YZ are in agreement with the findings of \cite{Sushch:2013tna}.

\begin{figure}[t!]
  \includegraphics[width=0.5\textwidth]{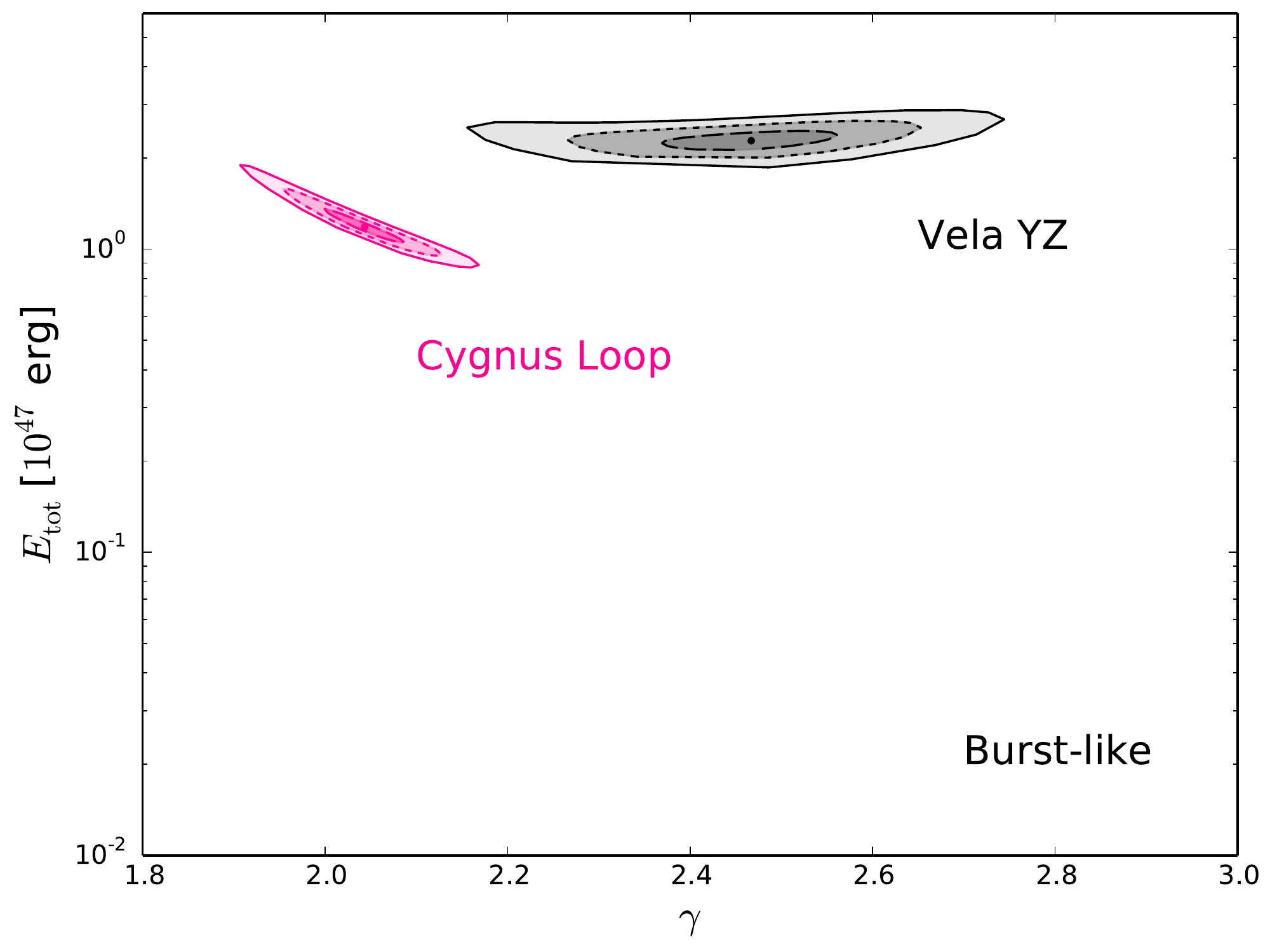}
  \includegraphics[width=0.5\textwidth]{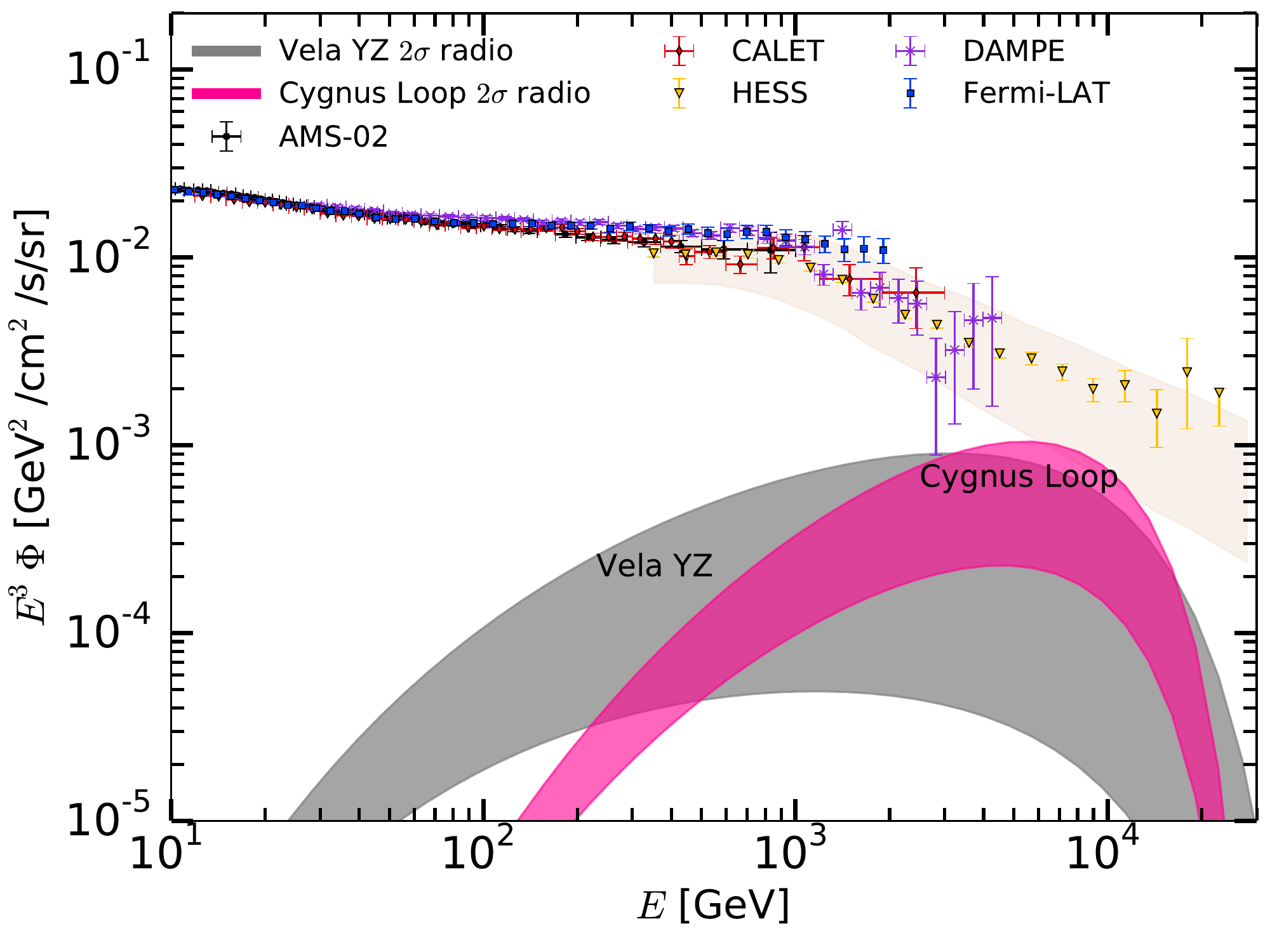} 
  \caption{Results of the fit to the radio spectrum for Vela YZ (gray) and Cygnus Loop (magenta). 
  Left: Regions of the parameter space  $E_{\rm{tot}}$, $\gamma$ selected by the fit to the radio spectrum. 
  The solid, dashed and long-dashed lines refer to respectively $3\sigma$, $2\sigma$ and $1\sigma$ contours for each source.
  Right: Prediction for the $ e^-$ flux from Vela YZ and Cygnus Loop using the values of $E_{\rm{tot}}$, $\gamma$
  within $2\sigma$ from the best fit to the radio spectrum shown in the left panel. 
  The $e^++e^-$ {\it Fermi}-LAT, AMS-02, DAMPE, HESS and CALET data with their statistics and systematic errors are also shown.
  }\label{fig:radioresults}
 \end{figure}

The  parameter space  $E_{\rm{tot}}$ -  $\gamma$ selected by the fit to the radio spectrum is reported in the left panel 
of Fig.~\ref{fig:radioresults} for both Vela YZ and Cygnus Loop, and for $3\sigma$, $2\sigma$ and $1\sigma$ confidence levels. 
This figure shows that radio data select narrow ranges for $\gamma$ and $E_{\rm{tot}}$.
For example, the $1\sigma$ contour for $\gamma_{\rm{Vela}}$ and $E_{\rm{tot, Vela}}$ is a few \% from the best fit.
Moreover, $E_{\rm{tot}}$ of the order of $10^{47}$ erg is in agreement with the usual expectations for the SNR energy budget, given the total energy released by a SN explosion in the ISM of  $\sim 10^{51}$~erg \cite{2007Sci...315..825M} and  a fraction conferred to $e^-$ of $\sim 10^{-5}-10^{-3}$ \cite{2009A&A...499..191T}.
We now evaluate the consequences of these results on the $e^++e^-$ flux. 
In the right panel of Fig.~\ref{fig:radioresults} we plot  the data on the $e^++e^-$ flux
along with the predictions for the flux from Vela YZ and Cygnus Loop  obtained by the parameters selected within the  $2\sigma$ contours in the left panel. 
The $e^++e^-$ flux data have not been used in this analysis and are displayed in the figure for illustrative purposes. 
The information in Fig.~\ref{fig:radioresults} is remarkable: the flux of $e^-$ from the closest SNRs as derived from a fit to radio data is slightly below the data on the inclusive flux.  
The flux from Vela YZ and Cygnus Loop can skim the HESS data, when all the uncertainties are considered.
In the assumption that all the radio emission is synchrotron radiation from $e^-$, our predictions indicate the highest flux expected from these sources
can shape the high energy tail of the $e^++e^-$ flux data.

 \begin{figure}[t]
  \includegraphics[width=0.5\textwidth]{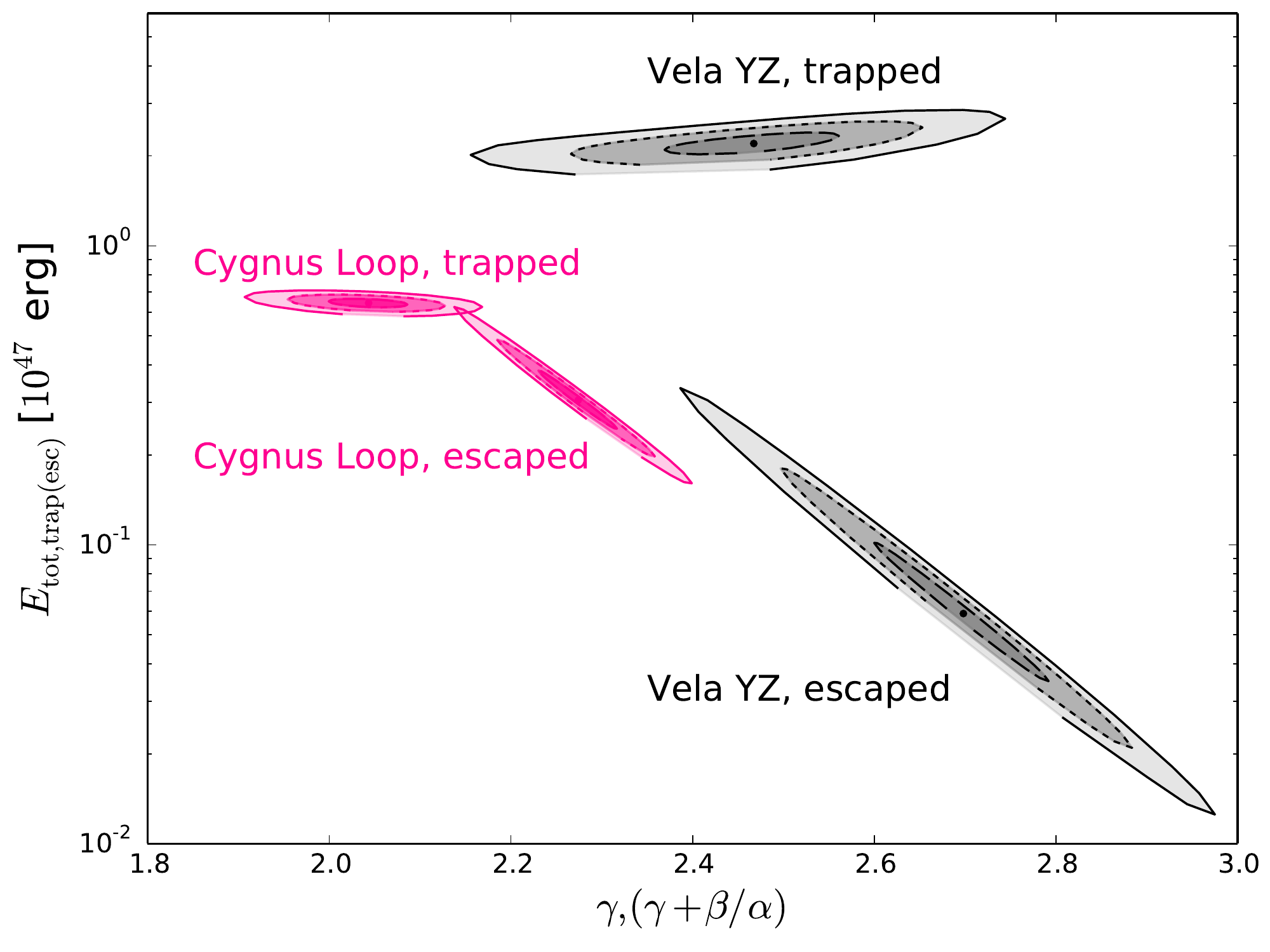}
  \includegraphics[width=0.5\textwidth]{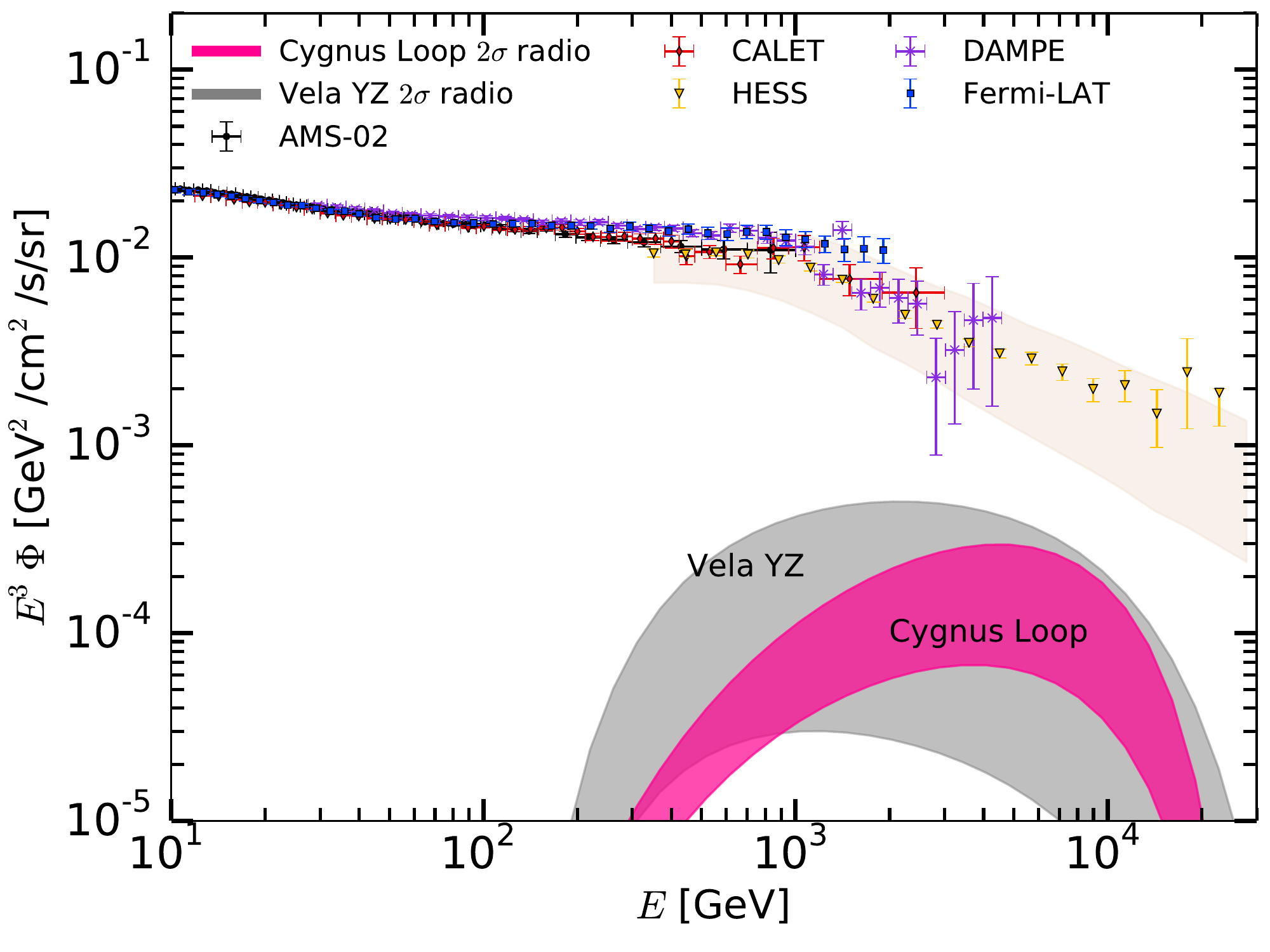} 
  \caption{\reply{Results of the fit to the radio spectrum for Vela YZ (gray) and Cygnus Loop (magenta) for the evolutionary model of the injection of $e^-$ from SNRs in Ref.~\cite{2012MNRAS.427...91O}. 
  Left: Regions of the parameter space  $E_{\rm{tot,trap}}$, $\gamma$ selected by the fit to the radio spectrum for Vela YZ (gray) and Cygnus Loop (magenta). 
  The derived regions for $E_{\rm{tot, esc}}$, $\gamma + \beta/\alpha$ are also reported for Vela YZ and Cygnus Loop. 
  The solid, dashed and long-dashed lines refer to respectively $3\sigma$, $2\sigma$ and $1\sigma$ contours for each source.
  Right: Prediction for the $ e^-$ flux from Vela YZ and Cygnus Loop using the values of $E_{\rm{tot, esc}}$, $\gamma+\beta/\alpha$
  within $2\sigma$ from the best fit to the radio spectrum shown in the left panel. 
  The $e^++e^-$ data are shown as in Fig.~\ref{fig:radioresults}, right panel. 
  }}\label{fig:radioresults_esc}
 \end{figure} 

\reply{We have explored the effects of the evolutionary escape model on the interpretation of radio spectrum for our two selected sources.   
In this case, the radio data are fitted through  Eq.~\ref{eq:Br} to tune the normalization and spectral index of the trapped $e^-$, namely the $Q_{0,\rm{trap}}$  and the index $\gamma$ of Eq.~\ref{eq:QTrap2}. The total energy of trapped $e^-$ is obtained as:
\begin{equation}
 E_{\rm tot, trap} = \int _{E_1} ^{ E_{\rm m, esc}(T)} dE \, E \,Q_{\rm trap}(E) \,,
 \label{eq:Etottrap}
\end{equation}
for each source. 
The normalization $A$ and spectral index of the escaped electrons are then derived by their relations with the trapped $e^-$, as derived in the evolutionary escape model in Ref.~\cite{2012MNRAS.427...91O}. 
By comparing Eq.~\ref{eq:QTrap2} and Eq.~\ref{eq:Qesc}, we obtain:
\begin{equation}\label{eq:relation_trap_esc}
 A=Q_{\rm 0,trap} E_{\rm knee}^{\beta/\alpha} \, \left( \frac{t_{\rm Sedov}}{T}\right)^{\beta}\quad, 
\end{equation}
while the spectral index of $e^-$ in Eq.~\ref{eq:Qesc} is simply $\gamma+\beta/\alpha$. 
The total energy of runaway $e^-$ for each source is then obtained as:
\begin{equation}
 E_{\rm tot, esc} = \int _{E_{\rm m, esc}(T)} ^{ \infty} dE \, E \,Q_{\rm esc}(E) \,.
 \label{eq:Etotesc}
\end{equation}
In Fig.~\ref{fig:radioresults_esc} (left panel) the parameter space $E_{\rm{tot, trap}}$ -  $\gamma$ ( $E_{\rm{tot, esc}}$ -  $\gamma +\beta/\alpha$) selected by the fit to the radio spectrum of Vela YZ and Cygnus Loop are shown for $3\sigma$, $2\sigma$ and $1\sigma$ confidence levels.
The selected intervals are narrow, and similar to the burst-like case for Vela YZ, see Fig.~\ref{fig:radioresults}. 
For Cygnus Loop the total energy of trapped $e^-$ is reduced by less than a factor of two. 
This is understood through  Eq.~\ref{eq:Etottrap}. The upper limit of the integral  is indeed $E_{\rm m, esc}(T)\sim 88~(17)$~GeV for Vela YZ (Cygnus Loop). 
The derived constraints on the parameter space $E_{\rm{tot, esc}}$ -  $\gamma + \beta/\alpha$ are also reported. 
Each of the two regions shows a strong correlation. 
The total energy  $E_{\rm tot, esc}$ of runaway $e^-$ is more than one orders of magnitude lower with respect to $E_{\rm tot, trap}$ for Vela YZ. As for Cygnus Loop, the difference is a factor of 2-3 for the best fit.  
This is again understood by the different $E_{\rm m, esc}(T)$ of the two sources. 
The  consequences of these results on the $e^++e^-$ flux are reported in Fig.~\ref{fig:radioresults_esc} (right panel). 
We plot again the data on the $e^++e^-$ flux,
along with the predictions for the flux of runaway $e^-$ from Vela YZ and Cygnus Loop. 
The flux of CR $e^-$ at Earth is computed  by using Eq.~\ref{eq:sol_esc}, and by using 
the parameters for the escaped $e^-$ selected within the  $2\sigma$ contours in the left panel. 
The flux from Vela YZ and Cygnus Loop is softer with respect to the burst-like approximation, reflecting the effect of the escape mechanism described in Eq.~\ref{eq:Qesc}. 
Moreover, compared to the burst-like approximation, the presence of an escape-limited maximum energy $E_{\rm m, esc}(T)$ for each source depletes the flux at Earth for $E<E_{\rm m, esc}(T)$. 
Considering all the uncertainties, under the evolutionary escape model the flux from Vela YZ and Cygnus Loop is predicted to contribute at most few percent to the data on the inclusive flux at TeV energies, a rough factor of two less than what is shown in Fig.~\ref{fig:radioresults}.
}

\section{\label{sec:flux}Results on the SNR  properties from $e^++e^-$ flux data}\label{sec:flux}
We now perform an analysis aimed at characterizing the $e^-$ emission from Vela YZ and Cygnus Loop SNRs through $e^++e^-$ (and $e^+$) flux data only. 
We want to assess the power of $e^++e^-$ data on the source properties with respect to the information brought by radio  (see Sect. \ref{sec:radio}) 
or the dipole anisotropy data (see Sect. \ref{sec:dipole}). We already know that these sources can contribute significantly to the  $e^-$ flux (see  \cite{DiMauro:2014iia,Manconi:2016byt,DiMauro:2017jpu}
and Fig.~\ref{fig:radioresults}). Therefore, we expect that the  $e^++e^-$ flux data will bound the contribution from local sources. This results will be quantified by 
bounds on  $\gamma$ and $E_{\rm{tot}}$, parameters effectively connected with the injection physics and the number of particles per unit energy released in the ISM.
\\
In order to explain the $e^++e^-$ data over many energy decades we consider, in addition to $e^{-}$ from SNRs, 
$e^{+}$ and $e^{-}$ produced by interactions of CRs on the ISM (secondary component) and by pair emission in PWNe \cite{Gaensler:2006ua, 2017SSRv..207..235B}.
We use here the model already employed in \cite{2010A&A...524A..51D,DiMauro:2014iia,Manconi:2016byt,DiMauro:2017jpu}. In particular, we refer to  \cite{Manconi:2016byt} 
for any detail.  We only outline here the main characteristics of the different contributors. 
The Galactic SNRs are divided into a {\it near} and a {\it far} population according to their distance $d$ from the Earth. As in Ref.  \cite{Manconi:2016byt}, we set $d = 0.7$ kpc. 
Since for sources near the Earth we dispose of abundant data, we model them individually picking their 
$d$, $T$, $B_r^{\nu}(\nu)$ and $\gamma$ from the Green's catalog \citep{Green:2014cea}. 
Far SNRs, for which the distance from Earth is $>0.7$ kpc, are instead assumed to be smoothly distributed in the Galaxy according to 
the spatial density profile in \cite{2015MNRAS.454.1517G}, and to inject $e^-$ in the ISM with an average  $E_{\rm{tot}}$ and $\gamma$.

As for the local SNRs, Vela YZ and Cygnus Loop will be modeled with free $E_{\rm tot}$ and spectral index $\gamma$. 
Instead, the contribution from Vela Jr is fixed to the leptonic model of \reply{\cite{2011ApJ...740L..51T}}. 
In particular we choose the values of \reply{$d=0.750$~kpc, $t=3$~ky, $B=12 \mu$G and $\gamma_{\rm{VelaJr}}=2.15$} \cite{2011ApJ...740L..51T,2008ApJ...678L..35K} and we compute the $Q_{0,\rm{VelaJr}}$ by using Eq.~\ref{eq:Br}. 
This choice is motivated by very limited information available for its radio flux, and because Vela Jr mainly contributes to $e^++e^-$ flux above 10 TeV where the very few data points are not constraining.

Similarly to SNRs, the injection spectrum $Q_{ \rm PWN}(E)$ of $e^{-}$ and $e^{+}$ emitted by a PWN  can be described by a power law with an exponential cut-off. The normalization of the PWN spectrum $Q_{0, \rm PWN}$ can be connected to the  spin-down energy of the pulsar $W_0$ by:
\begin{equation}
 \int_{E_{\rm min}}^{\infty}\,dE\,E\,Q_{ \rm PWN}(E)\,=\eta_{\rm PWN}\,W_0.
 \label{eq:PWN}
 \end{equation} 
 $W_0$ can be thus  constrained from the measured pulsar properties and assuming that the whole energy lost is carried by the magnetic dipole radiation \cite{2010A&A...524A..51D}. The factor $\eta_{\rm PWN}$ represents the efficiency with which the spin-down energy of the pulsar is converted into $e^{-}$ and $e^{+}$ pairs, 
and is expected to be of few $\%$ level~\cite{DiMauro:2014iia,DiMauro:2015jxa,Abeysekara:2017old}.
Our PWN sample is taken, as in~\cite{DiMauro:2014iia,DiMauro:2015jxa,Manconi:2016byt}, from the ATNF catalog \cite{ATNFcat}, from which we extract the spin-down energy, age and distance of each known PWN. All PWNe share a  common efficiency $\eta_{\rm PWN}$ and spectral index $\gamma_{\rm PWN}$, that will enter in our fits to the  $e^+$ and $e^-$ data as free parameters. Since the release of accelerated $e^-$ and $e^+$ pairs in the ISM  is estimated to occur after $40-50$ kyr after the pulsar birth \cite{2011ASSP...21..624B}, we select only sources with $t_{\rm obs}>50$~kyr.
Secondary leptons originated by the scatterings of proton and helium CRs off the ISM are modeled here following \cite{DiMauro:2015jxa}, using a free overall re-normalization factor $q$, which accounts for  uncertainties in the flux of primary CRs and in the production cross sections.

We fit the $e^++e^-$ flux data from HESS, CALET, DAMPE, AMS-02 and {\it Fermi}-LAT and AMS-02  $e^+$ flux with all the components described above.
We avoid strong biases from the solar modulation of the fluxes considering AMS-02, CALET and {\it Fermi}-LAT $e^{+}+e^{-}$ and AMS-02 $e^{+}$ data at $E>10$~GeV. 
We nevertheless include its effect in the force field approximation with the Fisk potential $\phi$ treated as  a free parameter. We use different $\phi_i$ in the fit to AMS-02, {\it Fermi}-LAT and CALET data since they cover different periods.
We take into account in the fit both the statistical and systematic uncertainties. 
We also include the uncertainty in the absolute energy scale taking $1.3\%$ for DAMPE \cite{Ambrosi:2017wek}, $5\%$ for CALET \cite{PhysRevLett.119.181101}, $0\%$ at 10~GeV to $5\%$ at 1 TeV with a linear trend in $\log{E}$ for {\it Fermi}-LAT, $2\%$ for $E=[10,290]$~GeV and $5\%$ for AMS-02, while for HESS we use the systematic band as reported in \cite{HESSICRC17}.
The fit is performed on the $e^++e^-$ and $e^+$ flux data with free parameters: $E_{\rm{tot},\rm{Vela}}$, $E_{\rm{tot},\rm{Cygnus}}$,  $\gamma_{\rm{Vela}}$, 
 $\gamma_{\rm{Cygnus}}$, $E_{\rm{tot}}$, $\gamma$, $\eta_{\rm PWN}$, $\gamma_{\rm PWN}$, $q$, $\phi_i$.
We only impose priors on $\gamma_{\rm{Vela}}$ ([1.90-3.10]) and   $\gamma_{\rm{Cygnus}}$ ([1.50-2.50]). 
The best fit ($\chi_{\rm red}^2 = \chi^2/\rm{d.o.f.} = 0.5$) parameters are 
 $\gamma_{\rm{Vela}}= 2.9\pm0.1$, $E_{\rm{tot},\rm{Vela}}=(2.4\pm0.2)\cdot10^{50}$~erg, 
 $\gamma = 2.70\pm0.06$, $E_{\rm{tot}}=(4.89\pm0.13)\cdot10^{47}$~erg,
$\eta_{\rm{PWN}}= 0.056\pm0.006$ and $\gamma_{\rm{PWN}} = 1.80\pm0.04$ for Vela YZ and Cygnus Loop SNRs.
 \begin{figure}[t]
  \centering
  \includegraphics[width=0.6\textwidth]{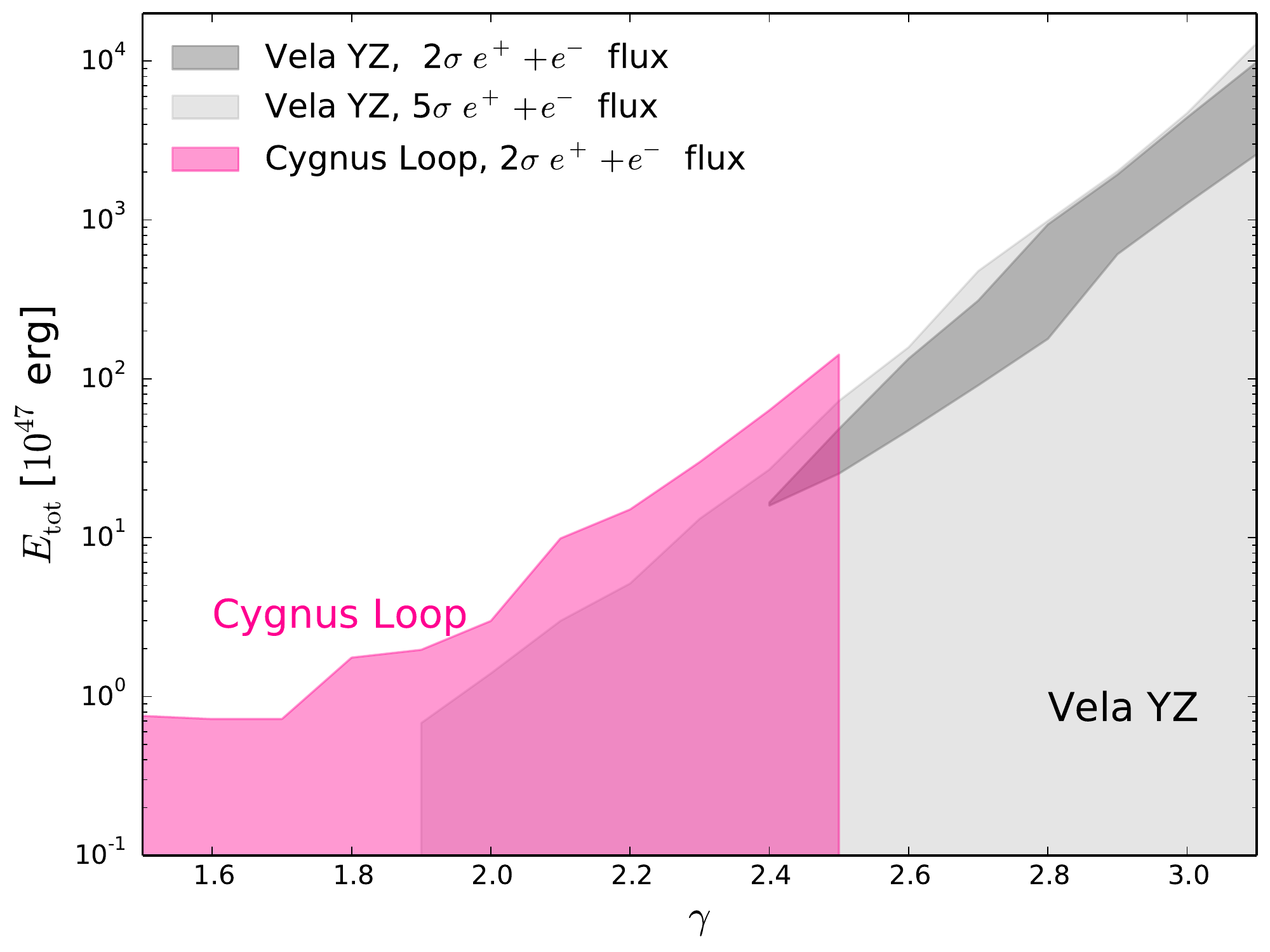}
  \caption{Regions of the parameter space $E_{\rm{tot}}$ -  $\gamma$ selected by the fit to the $e^{+}+e^{-}$ and $e^{+}$ flux data for  Vela YZ and Cygnus Loop. 
  The shaded regions denote the $E_{\rm{tot}}$, $\gamma$ values at a given number of $\sigma$ from the best fit for each source. The magenta region is for Cygnus Loop and $2\sigma$, while the 
   gray (light gray) regions are for Vela YZ and $2\sigma$  ($5\sigma$). 
  }\label{fig:fluxresults}
 \end{figure} 
The configurations within $2\sigma$ from the best fit for Vela YZ and Cygnus Loop free parameters are reported in Fig. \ref{fig:fluxresults}. 
For Vela YZ we also report the $5\sigma$ region which, as for Cygnus Loop, opens indeed to an upper bound. 
In the case of Vela YZ, we find that the 2$\sigma$ region from the best fit for $E_{\rm{tot},\rm{Vela}}$ and $\gamma_{\rm{Vela}}$ is narrow and the two parameters are strongly correlated.
The $E_{\rm{tot},\rm{Vela}}$ values selected by the $e^++e^-$ flux have no overlap (at $2\sigma$) with the ones constrained by the fit to radio flux data, and are systematically higher by at least one order of magnitude. The two regions fully overlap at $5 \sigma$ (see  Fig.~\ref{fig:radioresults}). 
If we perform the same analysis for DAMPE data alone, or for the combination of AMS-02, CALET and HESS data, the two regions fully overlap at $2 \sigma$.

\section{\label{sec:dipole}Results on the SNR  properties from $e^++e^-$ dipole anisotropy data}
We now assess the power of the recent  {\it Fermi}-LAT data on the $e^++e^-$ dipole anisotropy  $\Delta_{e^++e^-}$  \cite{Abdollahi:2017kyf}. 
This measure has provided upper bounds on  $\Delta_{e^++e^-}$ as a function of energy from 50 GeV up to about 1 TeV. 
We compute the relevant single source dipole anisotropy for the sources in our model, following \cite{Manconi:2016byt}. 
We remind here that the $\Delta_{e^++e^-}$  from a single source $s$ is given by:
\begin{equation}
\label{eq:eleposdipole}
  \Delta(E)_{e^+ + e^-} = \frac{3 K(E)}{ c} \frac{2 d}{\lambda^2(E, E_{s})} \frac{\psi_{e^+ + e^-}^{s}(E)}{\psi_{e^+ + e^-}^{tot}(E)},
\end{equation}
where $d$ is the distance to the source, 
$\lambda(E, E_{s})$ is the propagation scale defined in Eq.~\ref{eq:lambda},
$\psi_{e^+ + e^-}^{s}(E)$ is the $e^+ + e^-$ number density produced by the source $s$, and $\psi_{e^+ + e^-}^{tot}(E)$ is the total $e^+ + e^-$ number density obtained from the contributions of all the sources, both from isotropic smooth populations and from directional single sources. 
This expression can be appropriately associated to a physical observable whenever the source $s$ can be considered as dominant. 
In case more than one source is considered, the total dipole anisotropy may be computed as  \cite{Manconi:2016byt}:
\begin{equation}
\label{eq:dipolesources}
 \Delta(n_{max}, E)= \frac{1}{\psi^{tot}(E)} \cdot \sum_i \frac{\mathbf{r}_i\cdot \mathbf{n}_{max}}{||\mathbf{r}_i||}\cdot \psi_i(E)\, \Delta_i(E).
\end{equation}
Here $\psi_i(E)$ is the  number density of $e^-$ and/or $e^+$ emitted from each source $i$, $\mathbf{r}_i$ is the 
 source position in the sky and 
$n_{max}$ is the direction of the maximum flux intensity. The term $\psi^{tot}(E)=\sum_i \psi_i(E)$ is the total ($e^-$ and/or $e^+$) number density and includes the contribution from the discrete as well as all the isotropic sources. 
The anisotropy from each single source is given by $\Delta_i = \frac{3 K(E)}{c} \frac{|\nabla \psi_i(E) |}{\psi_i(E)}$, where the gradient is performed with respect to each source position.  
 \begin{figure}[t]
  \centering
  \includegraphics[width=0.6\textwidth]{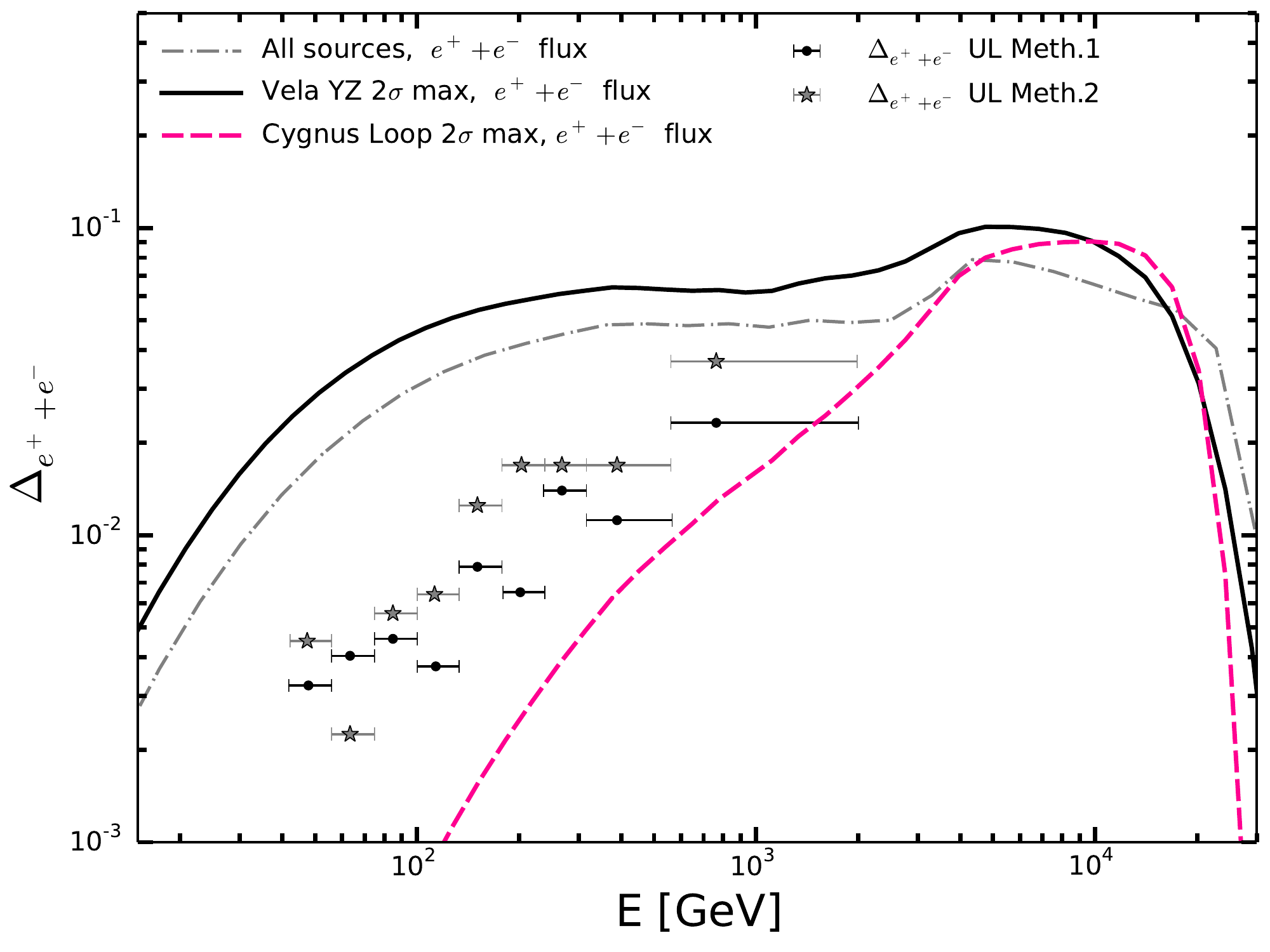}
  \caption{Dipole anisotropy predictions for Vela YZ and Cygnus Loop treated as single dominant sources (solid black and magenta lines, respectively), 
  and for all the sources combined together, shown  as gray dot-dashed line  (see text for details).
  The upper limits for {\it Fermi}-LAT dipole anisotropy are shown for the two different methods in \cite{Abdollahi:2017kyf}. 
  }\label{fig:dipoleresults}
 \end{figure} 
We compute the $\Delta_{e^++e^-}$ for Vela YZ and Cygnus Loop for all the parameters selected by the fit to ${e^++e^-}$ flux data \replyy{described in Sec.~\ref{sec:flux}} (at $2\sigma$ from the best fit), 
and reported in Fig. \ref{fig:fluxresults}. The maximum of  $\Delta_{e^++e^-}$ in each energy bin is then plotted as a black (magenta) solid line in Fig.~\ref{fig:dipoleresults} for Vela YZ (Cygnus Loop). 
We compare our predictions to the {\it Fermi}-LAT $\Delta_{e^++e^-}$ data (Bayesian Method 1 in \cite{Abdollahi:2017kyf}) 
above 100~GeV, to limit the effect from the solar wind \citep{Buesching:2008hr,STRAUSS20141015}. 
For Vela YZ, the anisotropy overshoots {\it Fermi}-LAT upper limits on the whole spectrum. We can therefore infer that {\it Fermi}-LAT data on the lepton dipole anisotropy add an independent piece of information
in addition to the flux data. This is one of the main results of this paper. The anisotropy amplitude data on {\it charged}
leptons have now the power to exclude configurations of the Vela YZ source spectrum, 
in principle compatible with the absolute flux data. 
For Cygnus Loop the conclusions are looser, since it shines at higher energies where the {\it Fermi}-LAT upper bounds are looser. 
In order to constrain Cygnus Loop parameters one would need dipole data at least up to 10 TeV. 

Since we are interested in the scenario in which the $\Delta_{e^++e^-}$ is maximal, we have checked for different effects that could lower our predictions.
In particular, we verified that also the total dipole anisotropy arising from all the individual sources entering in the predictions of the ${e^++e^-}$ and $e^+$ fluxes is not compatible with the experimental upper limits.
This is because Vela YZ is always the dominant contributor of the $e^-$ flux. 
We  computed the total anisotropy according to Eq.~\ref{eq:dipolesources} resulting from:
the local SNRs Vela YZ, Cygnus Loop and Vela Jr, and all the ATNF catalog PWNe.  The results shown in \replyy{Fig.  \ref{fig:dipoleresults} as a gray dot-dashed line} has been obtained
setting all the free parameters to their best fit to the ${e^++e^-}$ and $e^+$ fluxes data. The only prior being the flux data, Vela YZ turns out to dominate the flux as well as the dipole predicted at Earth. 
Moreover, we considered the potential effect of the guide magnetic field over the few hundred pc to the nearest sources, following what was done in \cite{2016PhRvL.117o1103A}. 
The local magnetic field properties were inferred by IBEX data ($l=210.5^{\circ}, b=-57.1^{\circ}$) \cite{0004-637X-776-1-30} from the study of the emission of high energy neutral atoms. 
As discussed in \cite{2016PhRvL.117o1103A}, the alignment of the dipole anisotropy of CRs with the total ordered magnetic field is demonstrated to potentially modify the phase of the observed CR dipole and lower its amplitude. 
We verified that projecting the dipole anisotropy of Vela YZ along the direction of the local magnetic field decreases the $\Delta_{e^++e^-}$  by a factor of roughly 2. 
Since the maximal Vela YZ anisotropy in Fig.~\ref{fig:dipoleresults} overshoots the {\it Fermi}-LAT upper limits by more than a factor of 3 up to $500$ GeV, also considering this effect would not change our conclusions.
Therefore, the dipole anisotropy in the CR lepton arrival direction sets additional tight constraints to the Vela YZ injection spectrum. 

 \begin{figure}[t]
  \centering
  \includegraphics[width=0.6\textwidth]{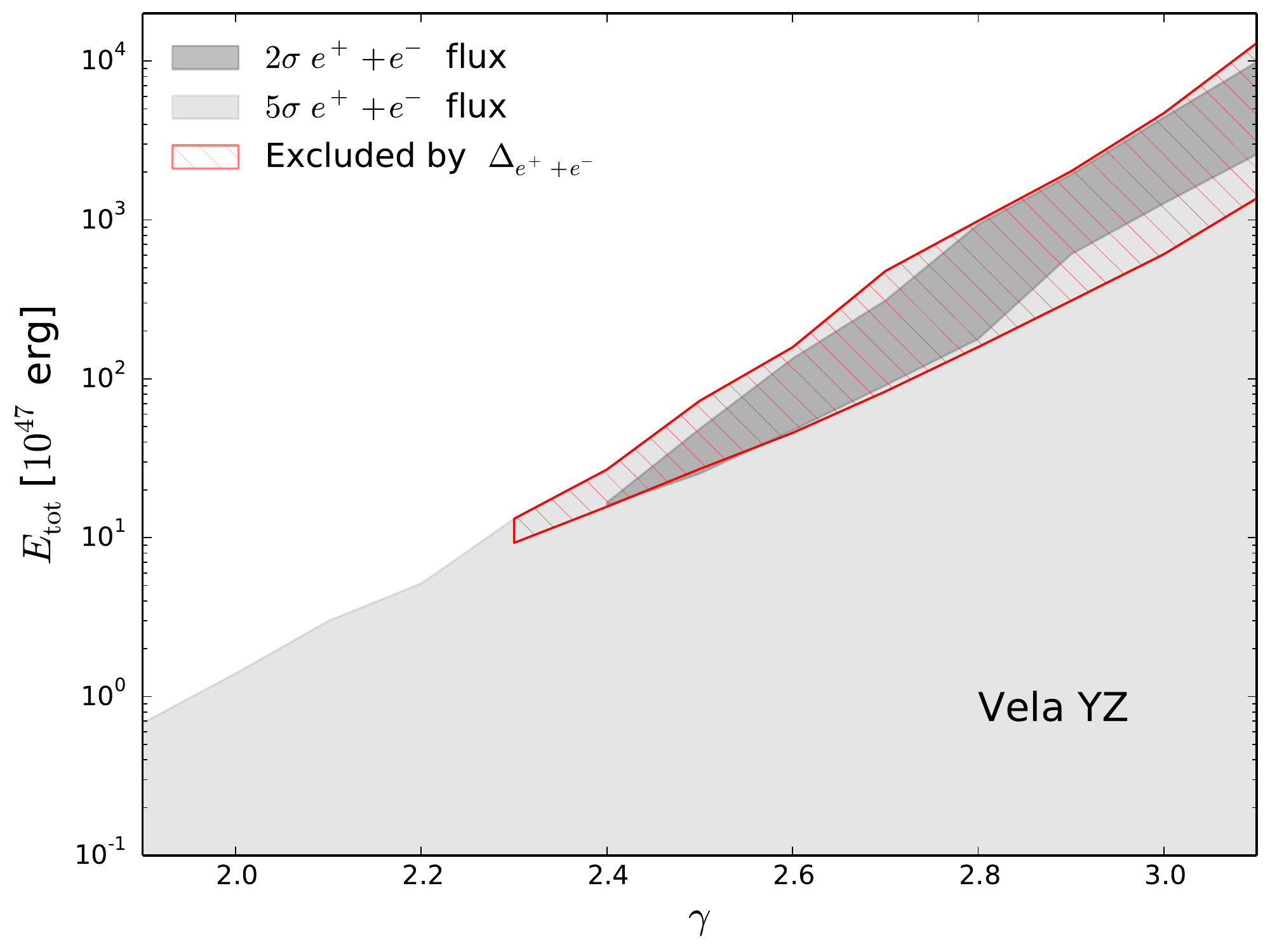}
  \caption{Dipole anisotropy constraints to the Vela YZ source parameters. 
  The regions of the parameter space $E_{\rm{tot}}$, $\gamma$ selected by the fit to the $e^{+}+e^{-}$ and $e^{+}$ flux data for  Vela YZ  are reported with shaded regions as in Fig.\ref{fig:fluxresults}. 
  The hatched  region denotes the configurations selected by $e^{+}+e^{-}$ and $e^{+}$ flux data and excluded by {\it Fermi}-LAT dipole anisotropy upper limits (Meth. 1) at $E>100$~GeV. 
  }\label{fig:dipoleconstraints}
 \end{figure} 
We now quantify the power of the dipole anisotropy to exclude configurations in the Vela YZ source parameters, otherwise compatible with the  $e^{+}+e^{-}$ flux data. 
We compute $\Delta_{e^++e^-}$ for all the configurations selected by the fit to the flux \replyy{described in Sec.~\ref{sec:flux}}. 
Whenever our predictions overestimate one data point at $E>$~100~GeV, the $E_{\rm{tot},\rm{Vela}} - \gamma_{\rm{Vela}}$ pair is considered as excluded. 
Very similar results are obtained when requiring two or more non-consecutive data points to be below the predictions, 
or if we employ only the two highest energy data points.
The results are displayed by the hatched region in upper panel of  Fig.~\ref{fig:dipoleconstraints}. 
The dipole anisotropy upper limits are not compatible  with the configurations selected by the fit to the flux data at 2$\sigma$, and with a subset of the configurations at 5$\sigma$.
Indeed, the anisotropy data exclude higher values of  $\gamma$, considered unlikely in acceleration models. 
The {\it Fermi}-LAT data on $\Delta_{e^++e^-}$ supplement a valuable information of the 
properties of Vela YZ, acting as a further physical observable for the understanding of the injection of $e^-$ in the ISM.

 \begin{figure}[]
  \includegraphics[width=0.5\textwidth]{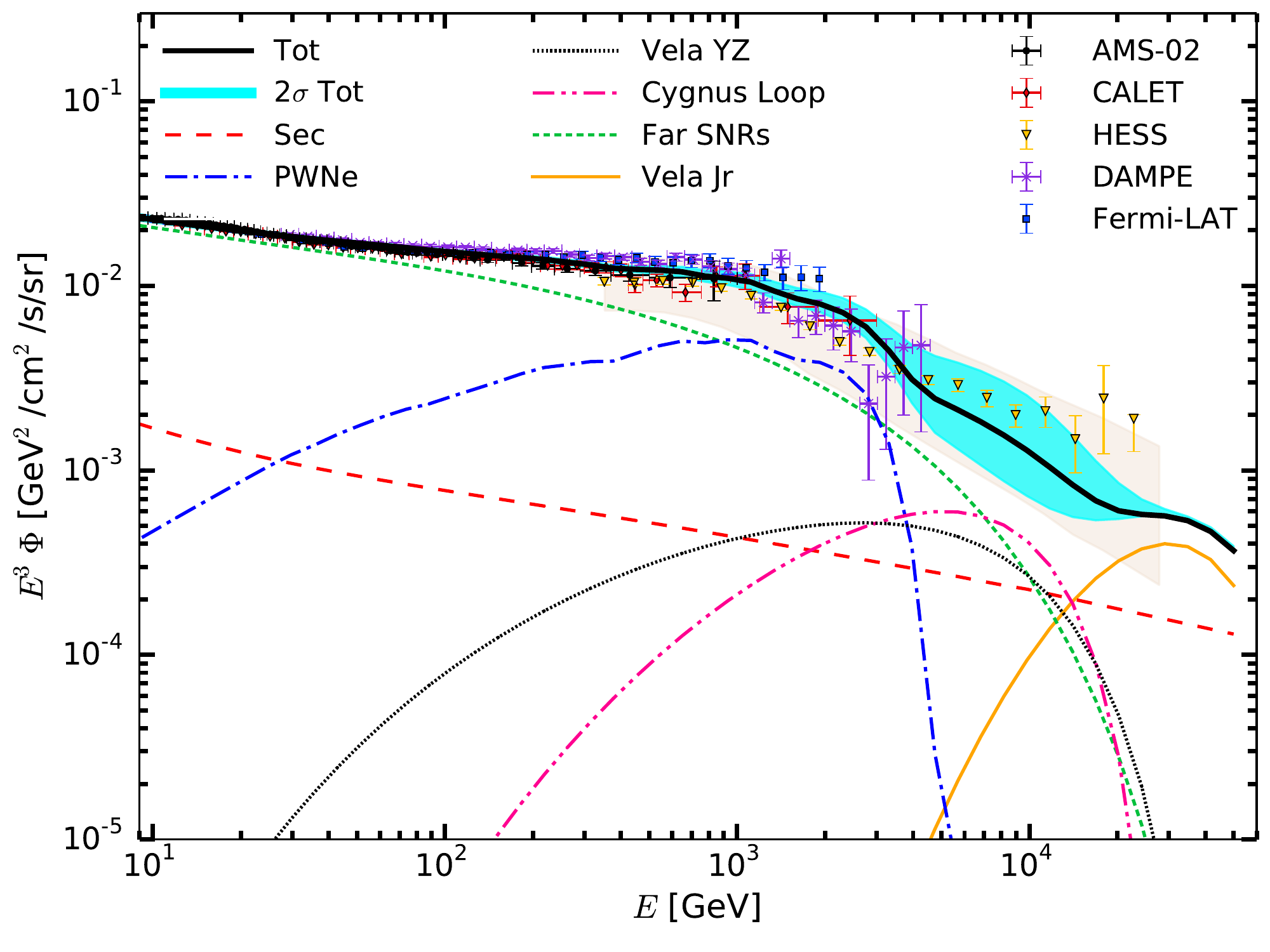}
   \includegraphics[width=0.5\textwidth]{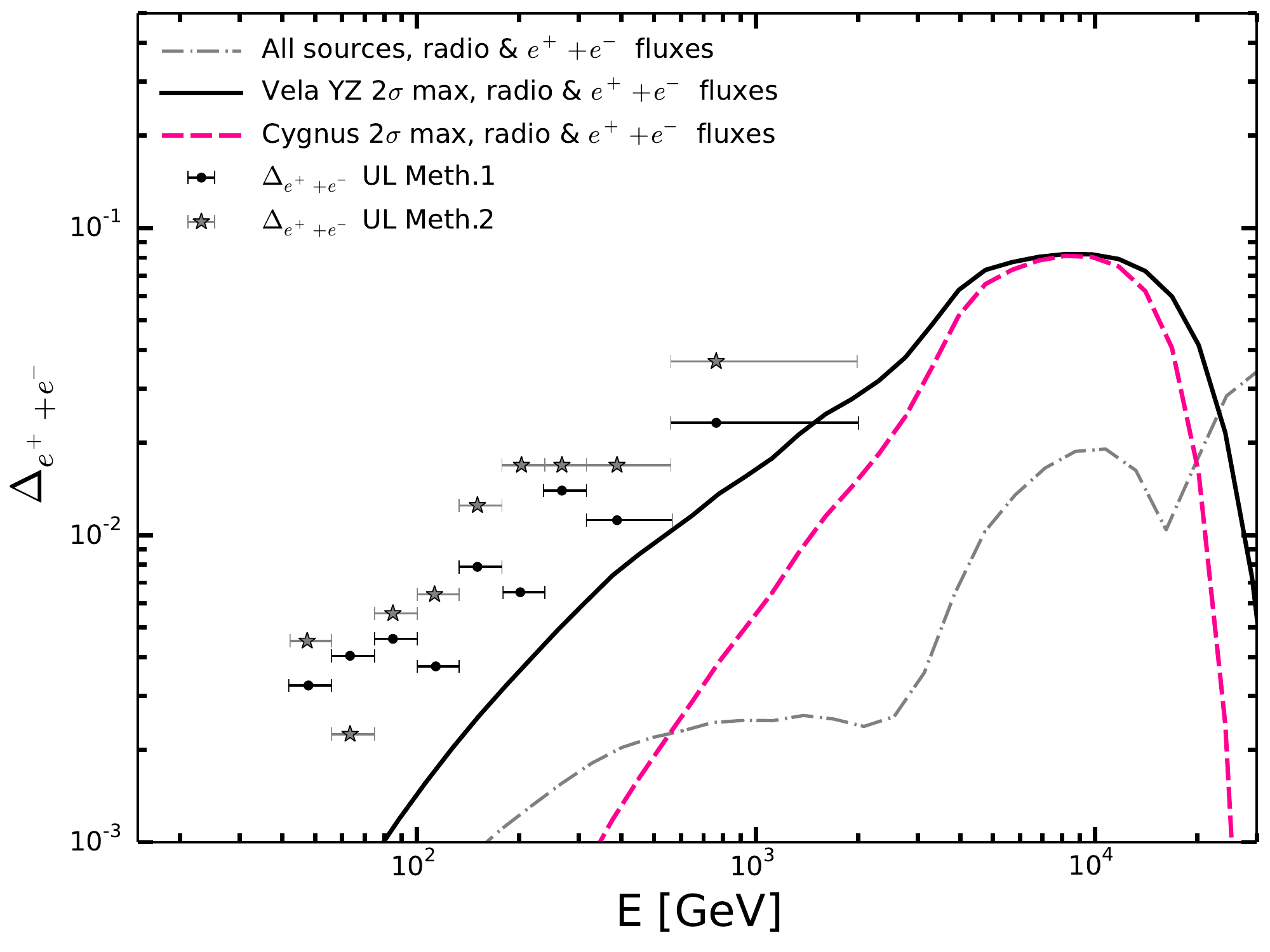}
   \caption{Results on the  $e^++e^-$ flux (left) and on the corresponding dipole anisotropies (right) from the multi-wavelength fit to all the data. 
   Left: The contribution from secondary production (red dashed), PWNe (blue dot dashed), Vela YZ (black dotted), Cygnus Loop (magenta dot-dot dashed), Vela Jr (orange solid) and the far smooth distribution of SNRs (green dotted) are shown. The $e^++e^-$ {\it Fermi}-LAT, AMS-02, DAMPE, HESS and CALET data with their statistics and systematic errors are also shown.
   Right: The maximal dipole anisotropy predicted for Vela YZ and Cygnus Loop as single dominant sources are reported with black solid and magenta dashed lines as in Fig.\ref{fig:dipoleresults}. 
  The total anisotropy resulting from the distribution of all the sources is shown with gray dot-dashed line.
  The upper limits for {\it Fermi}-LAT dipole anisotropy are shown for the two different analysis methods in \cite{Abdollahi:2017kyf}.
   }  
   \label{fig:mw}
 \end{figure} 

\section{\label{sec:multiw}Results  from multi-wavelength analysis}
We now combine all the three observables explored in the previous sections. 
Specifically, we compare the dipole anisotropy of Vela YZ and Cygnus Loop with the {\it Fermi}-LAT upper bounds, for the parameters of these sources selected by radio and $e^++e^-$ fluxes. 
We perform new fits on the  $e^++e^-$ and $e^+$ fluxes including the constraints for $E_{\rm{tot},\rm{Vela}}$, $\gamma_{\rm{Vela}}$, $E_{\rm{tot},\rm{Cygnus}}$, and $\gamma_{\rm{Cygnus}}$ derived from the fit to radio data. We minimize according to the following definition of the $\chi^2$:
\begin{equation}
\chi^2 = \sum^N_i \left(\frac{ \Phi^{\rm{model}}_i - \Phi^{\rm{data}}_i }{ \sigma^{\rm{data}}_i }\right)^2 + \sum^4_j \left(\frac{ \mathcal{P}^{\rm{model}}_j - \mathcal{P}^{\rm{data}}_i }{ \sigma^{\rm{data}}_{\mathcal{P},j} }\right)^2
\label{eq:chi}
\end{equation}
where the first term  is the statistical term that takes into account the difference between the model $\Phi^{\rm{model}}$  and the $e^+ + e^-$ flux data at $1 \sigma$ ($\Phi^{\rm{data}}$ and $\sigma^{\rm{data}}$). 
The second term runs over Vela YZ and Cygnus Loop $E_{\rm{tot}}$ and $\gamma$ and accounts for the deviation of these parameters in the model $\mathcal{P}^{\rm{model}}$ with respect to the best fit and $1\sigma$ error ($\mathcal{P}^{\rm{data}}$ and $\sigma^{\rm{data}}_{\mathcal{P}}$) as derived above in the fit to radio data.

We find a very good agreement between $e^++e^-$ and radio data ($\chi_{\rm red}^2\approx0.70$) with $\gamma_{\rm{Vela}} = 2.39\pm0.15$, $E_{\rm{tot}, \rm Vela} = (2.3\cdot\pm0.2)\cdot 10^{47}$ erg, $\gamma_{\rm{Cygnus}} = 2.03\pm0.05$ and $E_{\rm{tot}, \rm Cygnus} = (1.25 \pm0.06)\cdot 10^{47}$ erg for K15 propagation models. Using G15 propagation model the best fit parameters are extremely similar.
We illustrate in Fig.~\ref{fig:mw} the result of the best fit for all the components to the $e^++e^-$ flux.
We checked that all the predictions for the dipole anisotropy within $2\sigma$ from the best fit are below the {\it Fermi}-LAT upper bounds, as explicitly shown in  Fig.~\ref{fig:mw}.
The $\gamma$ for the spatially smooth distribution of SNRs is $2.48/2.44$ for K15/G15, respectively. 
We test different values for the cutoff energy of the smooth distribution and single SNRs. We find that the $\chi^2$ profile as a function of the cutoff energy is flat for $>10$ TeV while it worsens at low energy. The $95\%$ lower limit is at 8 TeV.
The putative $e^-$ injected by a radio unconstrained Vela SNR (see Sec.\ref{sec:flux}) are compensated in our framework
by the combination of $e^-$ produced by the Galactic smooth distribution of SNRs and all the PWNe.

 \begin{figure}[]
  \includegraphics[width=0.5\textwidth]{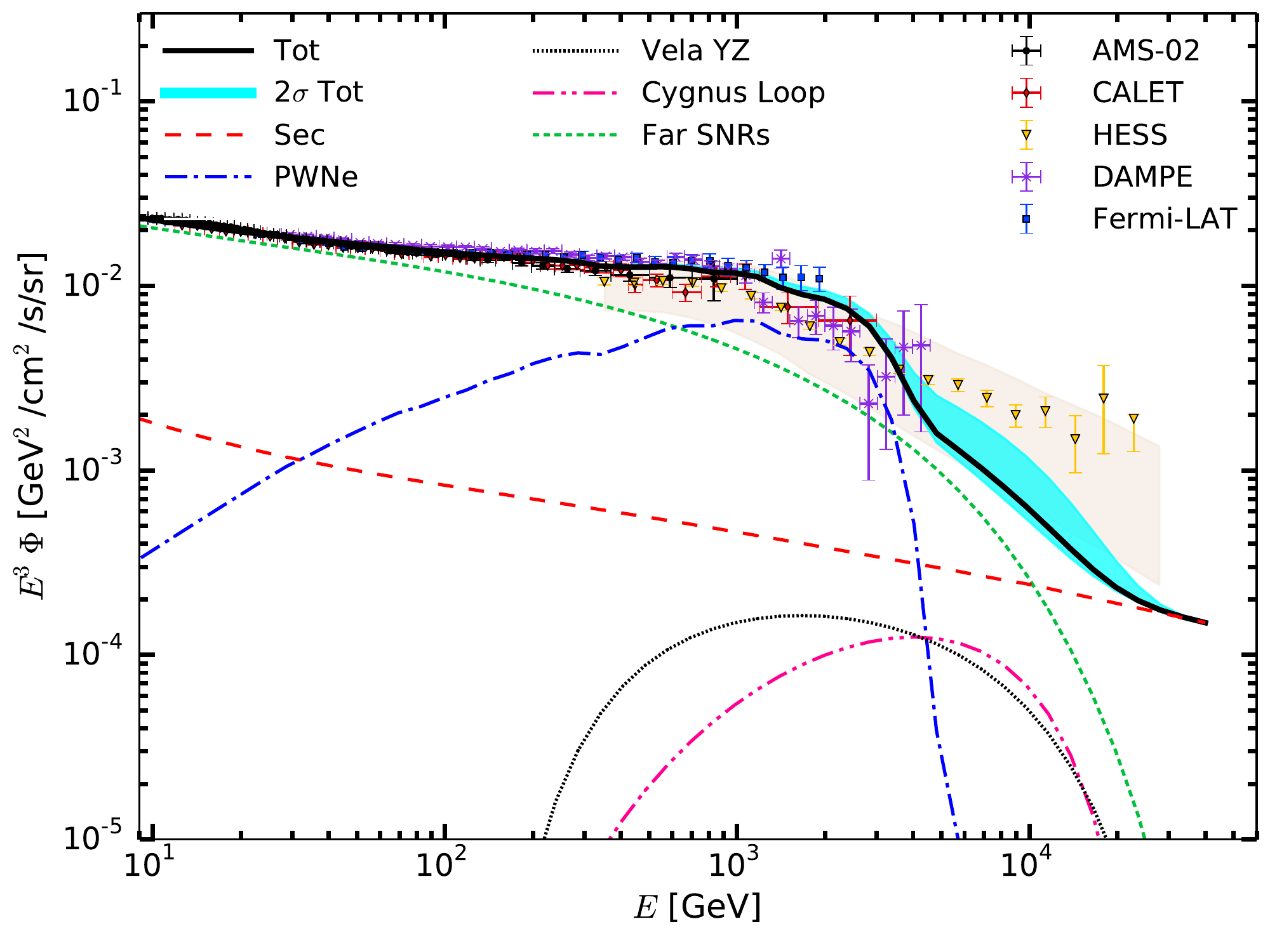}
   \includegraphics[width=0.5\textwidth]{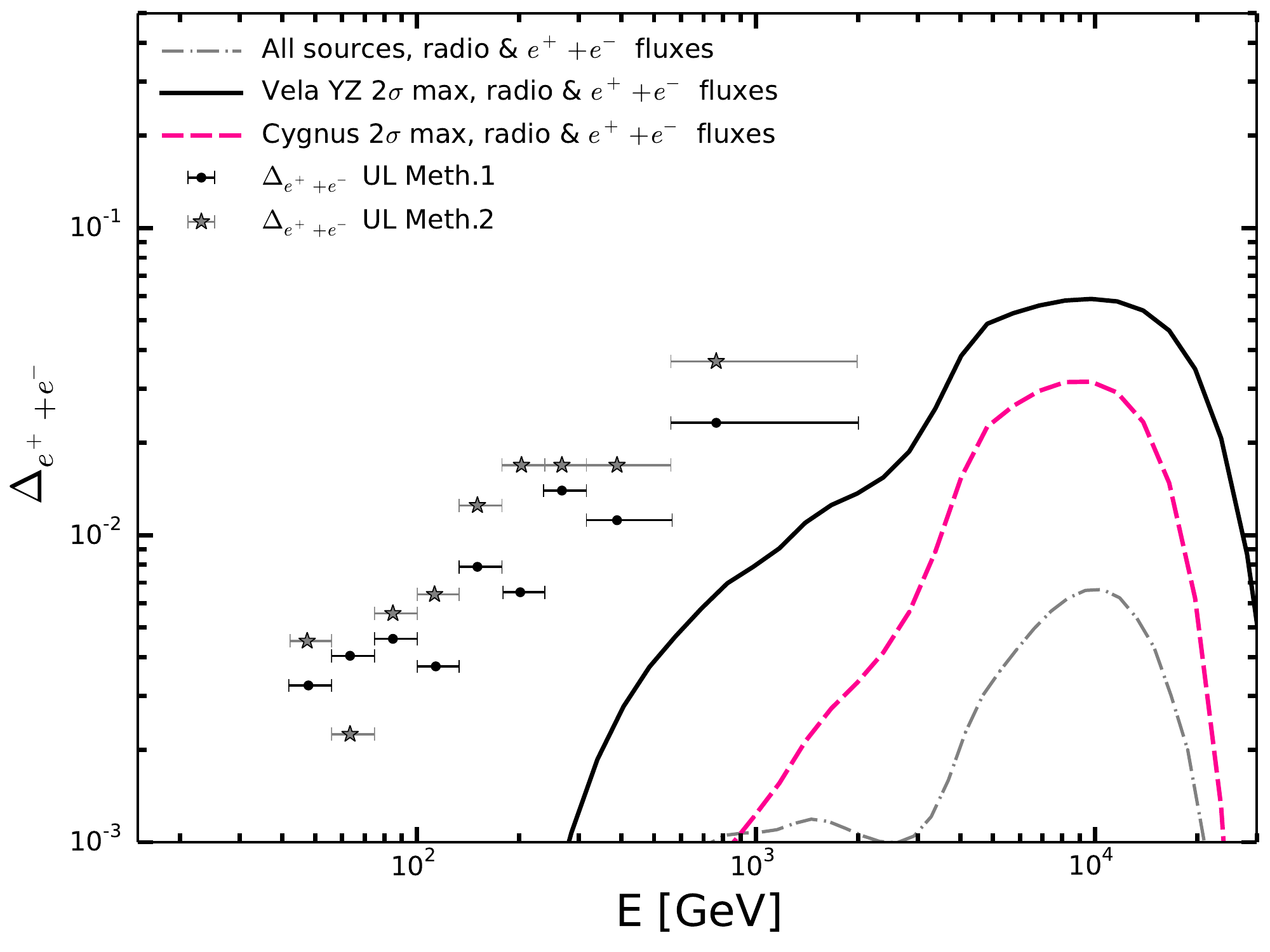}
   \caption{Same as Fig.~\ref{fig:mw} but using the evolutionary model of Ref.~\cite{2012MNRAS.427...91O} for the injection of $e^-$ by SNRs.}  
   \label{fig:mw_evol}
 \end{figure} 

\reply{In Fig.~\ref{fig:mw_evol} we report, on the same foot as Fig.~\ref{fig:mw}, the results obtained within the evolutionary escape model as discussed in Sec.~\ref{sec:model}.  
We find again a good fit ($\chi_{\rm red}^2\sim 0.87$), with  $\gamma_{\rm{Vela} + \beta/\alpha} = 2.66\pm 0.14$ and  $\gamma_{\rm{Cygnus}} = 2.27\pm0.06$  for K15 propagation model. The parameters which describes the smooth SNRs, the PWNe and the secondary component are compatible within the errors with respect to the burst-like scenario. 
Vela Jr has not been included in this analysis, since we found that in the evolutionary escape model its flux is suppressed, given its young age. 
We notice that the fluxes from Vela YZ and Cygnus Loop are now both below the secondary component, which instead is almost unchanged.  
This implies a slight increase of the PWNe contribution. We remind that also in this case the $e^+$ contribution is controlled by the AMS-02 $e^+$ flux data.  
As illustrated in Fig.~\ref{fig:mw_evol} (right panel), also in this case all the predictions for the dipole anisotropy within $2\sigma$ from the best fit are below the {\it Fermi}-LAT upper bounds.
\replyy{In the multi-wavelength analysis, the Vela YZ and Cygnus Loop SNRs parameters within both the burst-like and evolutionary models are not 
 constrained by the anisotropy data.}
With respect to the burst-like scenario in Fig.~\ref{fig:mw}, the predicted dipole anisotropies are decreased by more than a factor of 2 for TeV energies. 
}
Remarkably, we find a model which is compatible with all the $e^++e^-$ flux data, the radio data for Vela YZ and Cygnus Loop, and with the anisotropy upper bounds. 

\section{\label{sec:conc}Conclusions}
This paper proposes a \reply{new} multi-wavelength and multi-messenger approach aimed at improving the description of the local SNRs which
can contribute significantly to the measured  $e^++e^-$  flux. The latter is now measured with unprecedented statistics over 
several decades in energy. 

We work here within a framework in which the leptons measured at Earth from GeV up to tens of TeV energies have a composite origin. 
Specifically, $e^-$ are injected in the ISM by SNRs, and a symmetric source of  $e^\pm$ is provided by PWNe. Additionally, a low energy, asymmetric contribution of $e^+$ and $+e^-$
arises from the spallations of CRs on the ISM. In the understanding of the $e^-$ flux data, local sources, those located few hundreds parsecs from the Earth, may play a crucial role. 
The single, local SNRs that are found to be among the main contributors to the $e^-$ flux at $>10$~TeV are Vela YZ and Cygnus Loop. 
For these two sources, we develop a dedicated analysis to the injection spectrum of accelerated $e^-$ in the ISM. 

\reply{The injection of $e^-$ by SNRs into the ISM is treated following the burst like approximation, as commonly assumed in the literature.
Moreover, we have implemented an evolutionary escape model for the $e^-$ injection, and for the first time we have investigated its consequences on both the synchrotron emission and on the propagated CRs measuread at the Earth.}

We investigate the {\it compatibility} of these models for the emission and propagation of $e^-$  and $e^+$ in the Galaxy using three physical observables: 
\begin{itemize}
\item the {\it radio flux} at all the available frequencies from  Vela YZ and Cygnus Loop SNRs, 
\item the {\it $e^++e^-$ flux} from five experiments from the GeV to tens of TeV energy, 
\item the {\it  $e^++e^-$ dipole anisotropy} upper limits from 50 GeV to about 1 TeV. 
\end{itemize}

\noindent
We find that the {\it radio flux} for these nearby SNRs strongly constraints the total energy and the spectral index of the emitted $e^-$. 
\reply{In the case of the evolutionary escape model, we derive constraints on the total energy and spectral index of both trapped and runaway $e^-$.}
\reply{As for the burst-like approximation}, the flux of $e^-$ from Vela YZ and Cygnus Loop as derived from a fit to radio data is slightly below the data on the inclusive flux. 
It can skim the HESS data, when all the uncertainties are considered.
In the assumption that all the radio emission is synchrotron radiation from $e^-$, our predictions indicate the highest flux expected from these sources
can shape the high energy tail of the $e^++e^-$ flux data. 
\reply{In the case of the evolutionary escape model, the flux of runaway $e^-$ from Vela YZ and Cygnus Loop is slightly lower, }
\replyy{and their contribution to the $e^++e^-$ flux data is subdominant with respect to the other model components.}

\noindent
We perform a radio-blind analysis by fitting only and all the most recent {\it $e^+ + e^-$ flux} data. The data select correlated values for the total energy and spectral index of Vela YZ,
 and to a less extent of Cygnus Loop. The results for Vela YZ are compatible with the radio analysis within errors considered at  $5\sigma$ confidence level. 
 
\noindent
\reply{As a further novelty}, we consider the  upper limits on {\it $e^+ + e^-$ dipole anisotropy } as an \reply{additional} observable, and assess its power in constraining the Vela YZ and Cygnus Loop source properties. 
\reply{This operation is performed at the cost of no new free parameters.}
We find that the anisotropy overshoots {\it Fermi}-LAT upper limits on the whole spectrum when the Vela SNR parameters are left free to fit the $e^+ + e^-$ flux data \replyy{within a burst-like scenario}. 
For Cygnus Loop the conclusions are weaker, since it shines at higher energies where the {\it Fermi}-LAT  upper bounds are looser.
The results are very similar when all the single sources considered in the analysis (SNRs and PWNe) contribute to the anisotropy, which is 
dominated by Vela YZ. 
\replyy{For the first time, we show the severe constraints imposed by the most recent data on the  $e^+ + e^-$ anisotropy, what opens the opportunity of 
describing the most promising local sources of $e^-$ with {\it charged} lepton CRs.}

 \noindent
 We finally perform a multi-wavelength multi-messenger analysis by fitting
 simultaneously the radio flux on Vela YZ and Cygnus Loop and the $e^+ + e^-$ flux, and checking the outputs against the  $e^+ + e^-$ dipole anisotropy data. 
Considering the proper systematic uncertainties on the energy scale of the different data sets, we can fit the $e^+ + e^-$ spectrum on many energy decades using these local SNRs, a smooth distribution of SNRs, PWNe and secondary production. 
\replyy{In this case, the Vela YZ and Cygnus Loop SNRs parameters within both the burst-like and evolutionary models are not 
 constrained by the anisotropy data. }
Remarkably, we find a model which is compatible with all the $e^++e^-$ flux data, the radio data for Vela YZ and Cygnus Loop, and with the anisotropy upper bounds.

\begin{acknowledgments}
\reply{We warmly thank Y. Ohira for useful discussions.} 
SM gratefully acknowledges support by the Academy of Science of Torino through the 
{\it Angiola Agostinelli Gili} scholarship and the KIPAC Institute at SLAC for the kind hospitality. 
MDM acknowledges support by the NASA {\it Fermi} Guest Investigator Program 2014 through the {\it Fermi} multi-year Large Program N. 81303 (P.I. E.~Charles) and by the NASA {\it Fermi} Guest Investigator Program 2016 through the {\it Fermi} one-year Program N. 91245 (P.I. M.~Di Mauro). 
This work is supported by the "Departments of Excellence 2018 -2022"  Grant  awarded  by  the  Italian  Ministry  of  Education, University and Research (MIUR) (L. 232/2016). 
\end{acknowledgments}

\bibliography{dipole_JCAP}

\end{document}